\documentclass[aps,prx,twocolumn,superscriptaddress,nofootinbib,notitlepage,longbibliography,floatfix]{revtex4-2}
\usepackage{graphicx}
\usepackage{amsmath}
\usepackage{amstext}
\usepackage{amssymb}
\usepackage{xfrac}
\usepackage[colorlinks,citecolor=blue]{hyperref}
\usepackage{graphicx}
\usepackage{amsmath}
\usepackage{amstext}    
\usepackage{amssymb}
\usepackage{amsfonts}
\usepackage{longtable,booktabs}
\usepackage{hyperref}
\usepackage{url}
\usepackage{subfigure}
\usepackage{dsfont}
\usepackage{amsbsy}
\usepackage{dcolumn}
\usepackage{amsthm}
\usepackage{bm}
\usepackage{esint}
\usepackage{multirow}
\usepackage{hyperref}
\usepackage{cleveref}
\usepackage{mathrsfs}
\usepackage{amsfonts}
\usepackage{amsbsy}
\usepackage{dcolumn}
\usepackage{bm}
\usepackage{multirow}
\usepackage{color}
\usepackage{extarrows}
\usepackage{datetime}
\usepackage{comment}
\usepackage[super]{nth}
\usepackage{booktabs}
\usepackage[page,toc,titletoc,title]{appendix}
\usepackage[normalem]{ulem}

\hypersetup{
    colorlinks=true,
    linkcolor=blue,
    filecolor=magenta,
    urlcolor=blue,
}

\newcommand{\comments}[1]{}

\begin{document}
\title{Spin Texture in Doped Mott Insulators with Spin-Orbit Coupling}
\author{Shuai A. Chen}
\affiliation{Institute for Advanced Study, Tsinghua University, Beijing 100084,
China}
\author{Zheng-Yu Weng}
\affiliation{Institute for Advanced Study, Tsinghua University, Beijing 100084,
China}
\author{Jan Zaanen}
\affiliation{Institute-Lorentz for Theoretical Physics, Leiden University, P.O. Box 950, The Netherlands}
\date{\today }
\begin{abstract}
A hole injected into a Mott insulator will gain an internal structure
as recently identified by exact numerics, which is characterized by
a nontrivial quantum number whose nature is of central importance
in understanding the Mott physics. In this work, we show that a spin
texture associated with such an internal degree of freedom can explicitly
manifest after the spin degeneracy is lifted by a \emph{weak} Rashba
spin-orbit coupling (SOC). It is described by an emergent angular
momentum $J_{z}=\pm3/2$ as shown by both exact diagonalization
and variational Monte Carlo calculations, which are in good
agreement with each other at a finite size. In particular, as the internal
structure such a spin texture is generally present in the hole composite
even at high excited energies, such that a corresponding texture in
momentum space, extending deep inside the Brillouin zone, can be directly
probed by the spin-polarized angle-resolved photoemission spectroscopy
(ARPES). This is in contrast to a Landau quasiparticle under the SOC,
in which the spin texture induced by SOC will not be protected once
the excited energy is larger than the weak SOC coupling strength, away from the Fermi
energy. We point out that the spin texture due to the SOC should be
monotonically enhanced with reducing spin-spin correlation length
in the superconducting/pseudogap phase at finite doping. A brief discussion
of a recent experiment of the spin-polarized ARPES will be made.
\end{abstract}
\maketitle

\tableofcontents{}

\section{Introduction}
\subsection{Background}
A Landau's quasiparticle with a definite momentum, charge, and spin-1/2
has played a fundamental role in a conventional Fermi liquid. Whether
a doped charge (hole) can still propagate like a Bloch wave, is important
in understanding the two-dimensional (2D) Mott insulator at finite
doping. The latter is widely believed to be intrinsically related
to the high-$T_{c}$ superconductor in the cuprate \cite{anderson1987theresonating,lee2006doping}.

For a single hole injected into a quantum spin antiferromagnetic (AF)
background, although earlier studies \cite{lee2006doping,shraiman1988mobile,schmitt1988spectral,kane1989motion,martinez1991spin,liu1991spectral}
have indicated a Landau-like quasiparticle behavior based on effective
approaches, very recently it has been revealed by the unbiased
exact diagonalization (ED) and density matrix renormalization group
(DMRG) numerical calculations \cite{zheng2018hidden,zhu2018intrinsic,zhu2015quasiparticle},
that the doped hole will generate a composite structure by twisting the spin background via a nontrivial backflow
in the $t$-$J$ model. Manifestly a net spin current will further show up due to the presence of a spin-1/2 partner around the hole, which gains a nontrivial angular momentum,
$L_{z}=\pm1$, or a new double degeneracy under an open
boundary lattice (OBC) with a $C_{4}$ rotational symmetry
\cite{zheng2018hidden}.

\begin{figure}[htb]
\centering 
\includegraphics[scale=0.38]{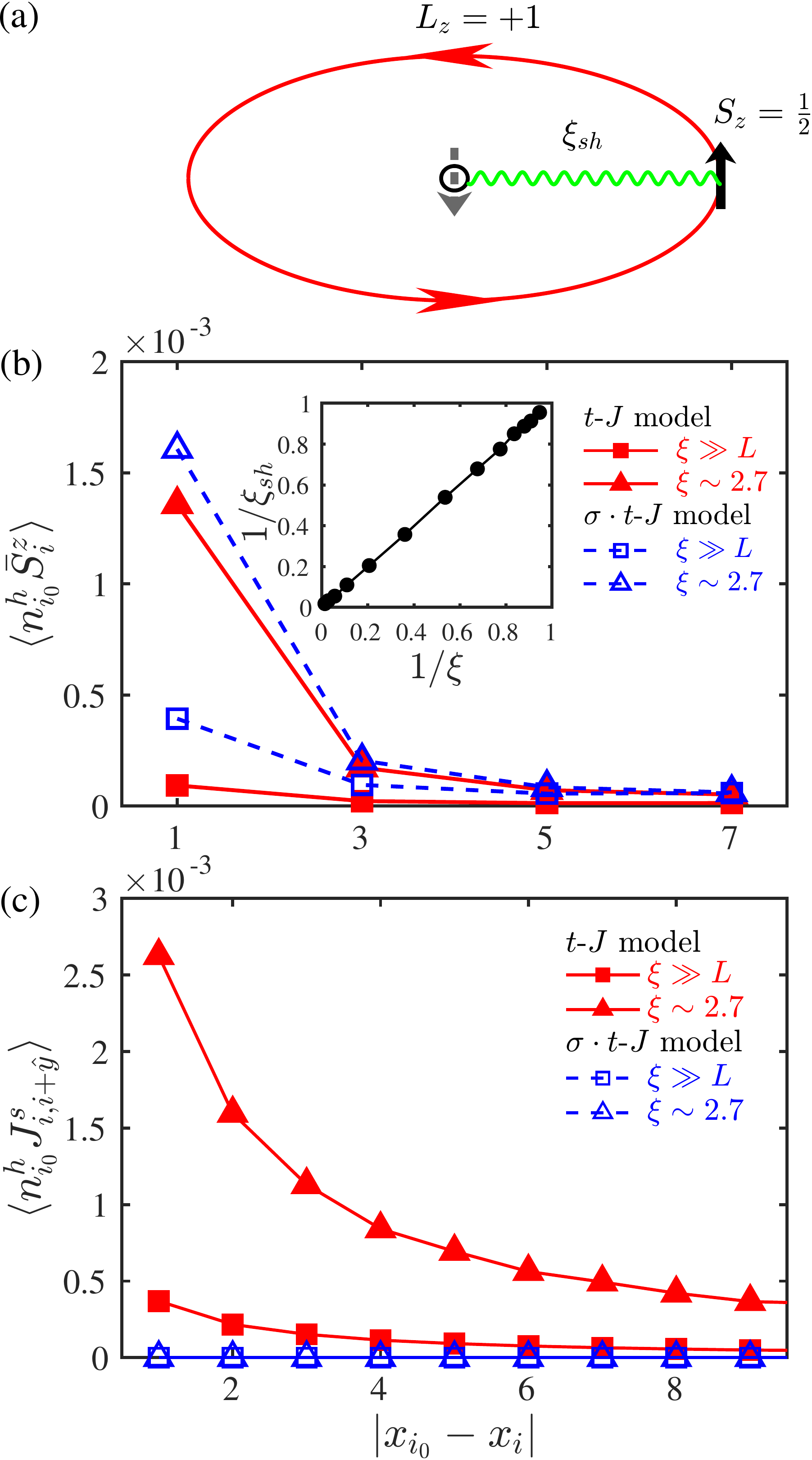} 
\caption{Composite structure of a doped hole in the pure $t$-$J$ model of a variational ground state description \cite{chen2019single}, which is schematically illustrated by (a): the hole (open circle) is associated with a net spin $S_{z}=\pm 1/2$ of a distance $\xi_{sh}$, which forms a circulating spin current around the hole. Such a hole composite acquires an angular momentum $L_{z}=\pm 1$ with novel ground state degeneracy in agreement with ED and DMRG \cite{zheng2018hidden};
(b) Here $\xi_{sh}$ is determined by the hole-spin correlator $\langle n_{i_0}^h\bar{S}_i^z\rangle$ in either the $t$-$J$  (red solid) or $\sigma\cdot t$-$J$ (blue dashed) model, which shows that $\xi_{sh}\simeq $ spin correlation length $\xi$ (the inset). Note that $\xi$ of the half-filling spin background is tunable, with two limits of long-range ($\xi\gg L$) and short-range ($\xi\sim 2.7 \ll L$) shown here (cf. Appendix \ref{APP:VMC}); 
(c) The spin current $J^{s}$, defined in Eq.~\eqref{Jsr}, around the hole as measured by the correlator $\langle n^h_{i_0}J^s_{i,i+\hat{y}}\rangle$  (along the $\hat{x}$-axis of the lattice) for the $t$-$J$ (red solid) model, which disappears in the $\sigma\cdot t$-$J$ (blue dashed) model. The VMC calculation is carried out in a finite-size square lattice of $N=L^2=20\times 20$.}
\label{Fig:schematicplot} 
\end{figure}
\begin{figure}[htb]
\centering 
\includegraphics[scale=0.5]{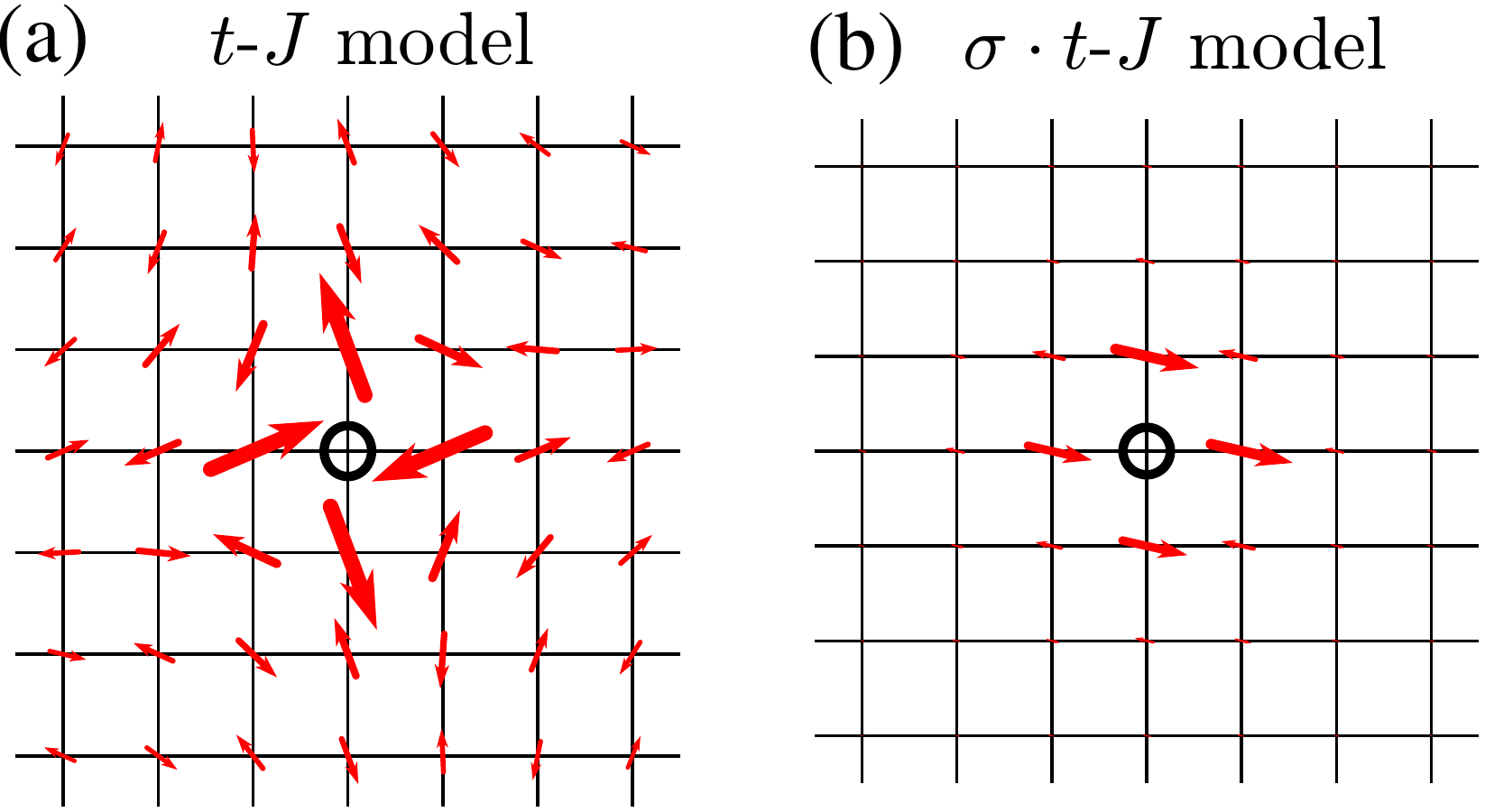} 
\caption{A spin texture pattern induced by a weak SOC, which is frozen into the x-y plane around a doped hole as characterized by a correlator $\langle n^h_{i_0}{\mathbf S}_i^{x,y}\rangle$ with $i_0$ located at the center marked by the open circle. (a) An explicit spin twist in the $t$-$J$ model; (b) The absence of the spin twist in the $\sigma\cdot t$-$J$ model. Here the half-filling spin background in the variational ground state study is chosen the same as in Fig. \ref{Fig:schematicplot} with $\xi\sim 2.7$ and the SOC strength $\lambda=0.01$ in units of $J=1$  (see text) . 
}
\label{Fig:spintexture} 
\end{figure}

Such a non-Landau-like novel structure of the composite hole revealed by ED and DMRG calculations can be well reproduced by
the variational Monte Carlo (VMC) study \cite{wang2015variational,chen2019single} of a variational ansatz state, 
which is illustrated in Fig.~\ref{Fig:schematicplot}. Generally a net spin-1/2 will be introduced to the ground state by a doped hole with a characteristic length scale $\xi_{sh}$. Taking the total $S_{z}=1/2$ without loss of generality, a net spin current around the hole is generally found as indicated in Fig.~\ref{Fig:schematicplot}(a). Namely there is an internal relative motion between the doped hole and the surrounding spins, which is characterized by an emergent quantum number, i.e., the angular momentum $L_{z}$ mentioned above. 

The detailed composite structure is further shown in Figs.~\ref{Fig:schematicplot}(b) and (c) based on the correlators, $\langle n^h_{i_0}\bar{S}^z_i\rangle$ and $\langle n^h_{i_0}J^s_{i,i+\hat{y}}\rangle$, which measure the distribution of the net $S_{z}=1/2$ around the doped hole [(b)] and the corresponding circulating spin current [(c)], respectively. Here $n^h_{i_0}$ denotes the hole number at site $i_0$, $\bar{S}^z_i$ the summation of two nearest-neighboring spins along the ${x}$-axis, i.e., $S^z_i+S^z_{i+\hat{x}}$, with $i$ in the opposite sublattice site of $i_0$ such that a staggered oscillation is smoothened, and $J^s_{i,i+\hat{y}}$ the transverse spin current [cf. Eq.~\eqref{Jsr}]. The variational ground state \cite{chen2019single} [cf. Eq.~(\ref{VMC1h})] is obtained at a lattice of $N=L^2=20\times 20$. The average distance of the $S_{z}=1/2$ from the hole, $\xi_{sh}$, is shown in the inset of Fig.~\ref{Fig:schematicplot}(b), which is essentially proportional to the spin-spin correlation length $\xi$ of the spin AF background that can be artificially tuned in the VMC calculation (two cases of $\xi\gg L$ and $\xi\sim2.7$ in units of the lattice constant are shown in the main panel with $L$ denoting the sample length). Then, Fig. \ref{Fig:schematicplot}(c) further shows that the spin current around the hole gets enhanced with reducing $\xi_{sh}\simeq \xi$. 
It is noted that here we are always focused on the one-hole case in a finite-size sample. If $L\rightarrow \infty$, a true AF long-range order $\xi\rightarrow \infty$ will set in, such that spin excitations become
gapless, which may also contribute to the spin currents induced by the hole in addition to just an $S^z=1/2$ shown in Fig.~\ref{Fig:schematicplot}(a) at the finite lattice size. We do not consider this singular limit in the following.

Note that there are two chiralities of the spin currents in the one-hole degenerate ground states corresponding to the novel quantum number $L_{z}=\pm1$ (only $L_{z}=1$ is shown here). 
By contrast, the chiral spin current can be turned off with vanishing angular momentum in the
so-called $\sigma\cdot$$t$-$J$ model, in which the doped hole restores the behavior of a Landau-like quasiparticle \cite{chen2019single}.
Here the sole distinction between the $t$-$J$ and $\sigma\cdot t$-$J$
models is characterized by the nontrivial backflow of spin current in the former, which is absent in the latter with the disappearance of the novelty of the hole composite, although $\xi_{sh}$ remains similar in both cases [cf. Fig. \ref{Fig:schematicplot}(b)].

It is then natural to raise the following question, i.e., how can one detect such an internal quantum structure of the composite hole experimentally? 
To address this question, in this paper, we introduce a \emph{weak} Rashba spin-orbit coupling (SOC), which may be present in a double-layer system of the cuprate materials like Bi$2212$ \cite{gotlib2018revealing}. Then we show that the hidden transverse spin current pattern appearing around the hole in Fig. \ref{Fig:schematicplot} can manifest explicitly an emergent semiclassical spin texture lying in the $x$-$y$ plane, as illustrated in Fig. \ref{Fig:spintexture}(a). Here a real space distribution of the correlator $\langle n^h_{i_0}{\mathbf S}_i^{x,y}\rangle$ is presented with the hole site $i_0$ fixed at the center of the sample (denoted by the open circle in Fig. \ref{Fig:spintexture}. Namely, the novel internal composite structure of the hole in the pure (without SOC) $t$-$J$ model can be effectively visualized via a frozen transverse spin texture, which is  induced by a weak SOC lifting the spin degeneracy and resulting in the locking of $S_{z}$ with $L_{z}$ by a total angular momentum $J_{z}=L_{z}+S_{z}$. 
By contrast, such a transverse spin texture is absent in the $\sigma\cdot t$-$J$ model in the same weak SOC [cf. Fig.\ref{Fig:spintexture}(b)]. In the latter, the single hole behaves like a Bloch wave without a surrounding spin current [cf. Fig. \ref{Fig:schematicplot}]. In the following, we further briefly summarize the main results obtained in the present work and their physical implications.

\subsection{Main results}
In this work, we show that the hidden spin current pattern of the spin-1/2 partner in a hole composite can manifest itself as an emergent semiclassical spin texture with the introduction of a weak Rashba SOC. It is explicitly exhibited in either the spatial space as indicated by Fig. \ref{Fig:spintexture}(a) above and the momentum space (see below) as found by ED as well as the VMC calculation.

As shown in Fig.~\ref{Fig:spintexture}(a), such a spin twist as measured by the correlator $\langle n^h_{i_0}{\mathbf S}_i^{x,y}\rangle$ is formed in the $x$-$y$ plane around the hole. 
Physically it reflects the underlying antiparallel spin alignments in the $x$-$y$ plane along either the ${x}$ or ${y}$ direction across the hole to facilitate the hopping. By a sharp contrast, no such a relative spin twist is shown in Fig.~\ref{Fig:spintexture}(b) for the $\sigma\cdot t$-$J$ model, which is consistent with the the absence of a spin current structure in Fig.~\ref{Fig:schematicplot}(c). Note that $\xi\sim2.7$ is chosen in the VMC calculation such that the distribution of the spin-1/2 partner is pulled closer to the hole to enhance the effect in Fig.~\ref{Fig:spintexture}.

The energy shifts of the ground state characterized by $J_{z}=\pm3/2$ and the first excited one at $J_{z}=\pm1/2$, together with the ground state energy shift in the $\sigma\cdot t$-$J$ model, are calculated by both ED and VMC methods as a function of the SOC strength $\lambda$, which are in excellent agreement as presented in Fig.~\ref{Fig:EnergyEDVMC}. Here it is interesting to note that the effect of the SOC in the $\sigma\cdot t$-$J$ model is much weaker than that in the ground state energy of the $t$-$J$ model, indicating that the spin current structure substantially enhances the effect of the Rashba SOC.   

Furthermore, under the rotational symmetry, a finite angular momentum $J_{z}$ in the $t$-$J$ model also implies a spin twist structure in momentum space, consistent with the Rashba SOC $\sim{\bf S}_{\bf k}\cdot({\bf k}\times\hat{{\bf z}})$ where ${\bf S}_{\bf k}$ denotes a single particle's spin at momentum ${\bf k}$. Indeed, in Fig.~\ref{Fig:helicity}, the spin textures of $\langle {\bf S}_{\bf k}\rangle$ lying in the $x$-$y$ plane, characterized by a spin
helicity in the momentum space, are shown for the ground state of $J_{z}=\pm3/2$ and the first excited state of $J_{z}=\pm1/2$, respectively, as calculated by ED. Again the spin helicity is much weaker for the $\sigma\cdot t$-$J$ model [cf. Fig.~\ref{Fig:helicity}(a)]. In Fig.~\ref{Fig:OBCvortex}, the ED and VMC results are further compared in consistency. Both ED and VMC calculations also indicate that the real-space spin texture in Fig.~\ref{Fig:spintexture}(a) can persist into high excited states.
At a higher excitation energy, the corresponding spin texture in momentum space will be deeper inside the Brillouin zone, as shown in Fig.~\ref{Fig:8by8} obtained by VMC at a larger sample size. Such a hole-like quasiparticle with a spin texture in momentum space may be thus directly probed by the spin-polarized angle-resolved photoemission spectroscopy (ARPES) measurement \cite{gotlib2018revealing}. 

The persistence of the spin texture in the Brillouin zone, even close to the $\Gamma$ point in Fig.~\ref{Fig:8by8}, is a fact that is in sharp contrast to a Landau Fermi liquid. In the latter, the spin
of a quasiparticle may also lock with its momentum to form a spin
helicity in the momentum space by SOC. But it is a single particle picture without the
\emph{spatial} spin texture [cf. Fig.~\ref{Fig:spintexture}(a)], in which the SOC-induced spin helicity is expected to be only robust
near the Fermi surface within a very narrow energy interval
comparable with the strength of the weak SOC. Away from the Fermi
energy, the spin helicity of a Landau quasiparticle can be easily washed away by scattering (cf. Fig.~\ref{Fig:SpinTextureFL}) 
without the protection from the many-body effect in the $t$-$J$ case as represented by a real-space
spin texture associated with $J_z\neq 0$.

Finally, we argue that even though this work has mostly focused on a single hole in a finite size system, where
ED and VMC calculations are available to compare with each other, one may reasonably generalize the main conclusion to the experimentally relevant regime at finite doping. Firstly, it has been shown by the VMC approach that the spin texture amplitude gets enhanced with reducing $\xi$ in the spin background (cf. the inset of Fig.~\ref{Fig:fintedoping}), which is due to the fact that the composite of a hole and its spin-1/2 partner gets smaller with an enlarged spin current.
In order to make quantitative comparison with the experiment \cite{gotlib2018revealing}, aside from the absolute amplitude of the spin-polarized spectral function in the inset, a relative polarization ratio of the spectral function with spin perpendicular to momentum is also shown in the main panel of Fig.~\ref{Fig:fintedoping} at a given momentum, which indeed indicates a comparable strength with the experimental observation \cite{gotlib2018revealing}. 
Secondly, we point out that the pairing of the doped holes with the spin texture structure obeying the time-reversal symmetry is presumably between the composite holes with the opposite pseudo-spin $J_{z}=\pm3/2$ in the presence of SOC. Those other quasiparticle excitations with distinct spin textures, which have higher energies
due to the SOC splitting, are thus effectively excluded out of the pair condensate.
In such a superconducting/pseudogap regime, predominantly the chiral spin texture associated with the
unpaired quasiparticles of $J_{z}=\pm3/2$  (which exhibit the \emph{same} spin texture in momentum space) will be observed
by the spin-polarized ARPES measurement over a finite range of energy,
which covers the extended momentum space away from the Fermi surface
region. 
Such a many-body picture clearly illustrates why a quasiparticle excitation with spin texture is ``protected'', which can survive even when the thermal energy in superconducting/pseudogap regime is dominant over the small energy scale of SOC. 
By contrast, the emergent Bogoliubove/Landau quasiparticle with Fermi arc \cite{zhang2020arpes} should only contribute to a negligible spin texture under a weak SOC. 

The rest of paper will be organized as follows. In Sec.~\ref{sec:Models}, the $t$-$J$ and $\sigma\cdot t$-$J$ models with a Rashba SOC are introduced. In Sec.~\ref{sec:1holestudy}, we focus on a single hole's motion in a finite-size lattice and discuss the spin texture structure induced by the SOC based on both ED and VMC approaches. In Sec.~\ref{sec:finitedoping}, a qualitative discussion at a finite doping is presented. Finally, Sec.~\ref{sec:conclusion} is devoted to the concluding remarks. Various technical details and analyses are presented in Appendices.

\section{Models}
\label{sec:Models}
\subsection{The $t$-$J$ and $\sigma\cdot t$-$J$ models}

\label{Sec::tj-spincur}

\begin{figure}[t]
\centering 
\includegraphics[scale=0.36]{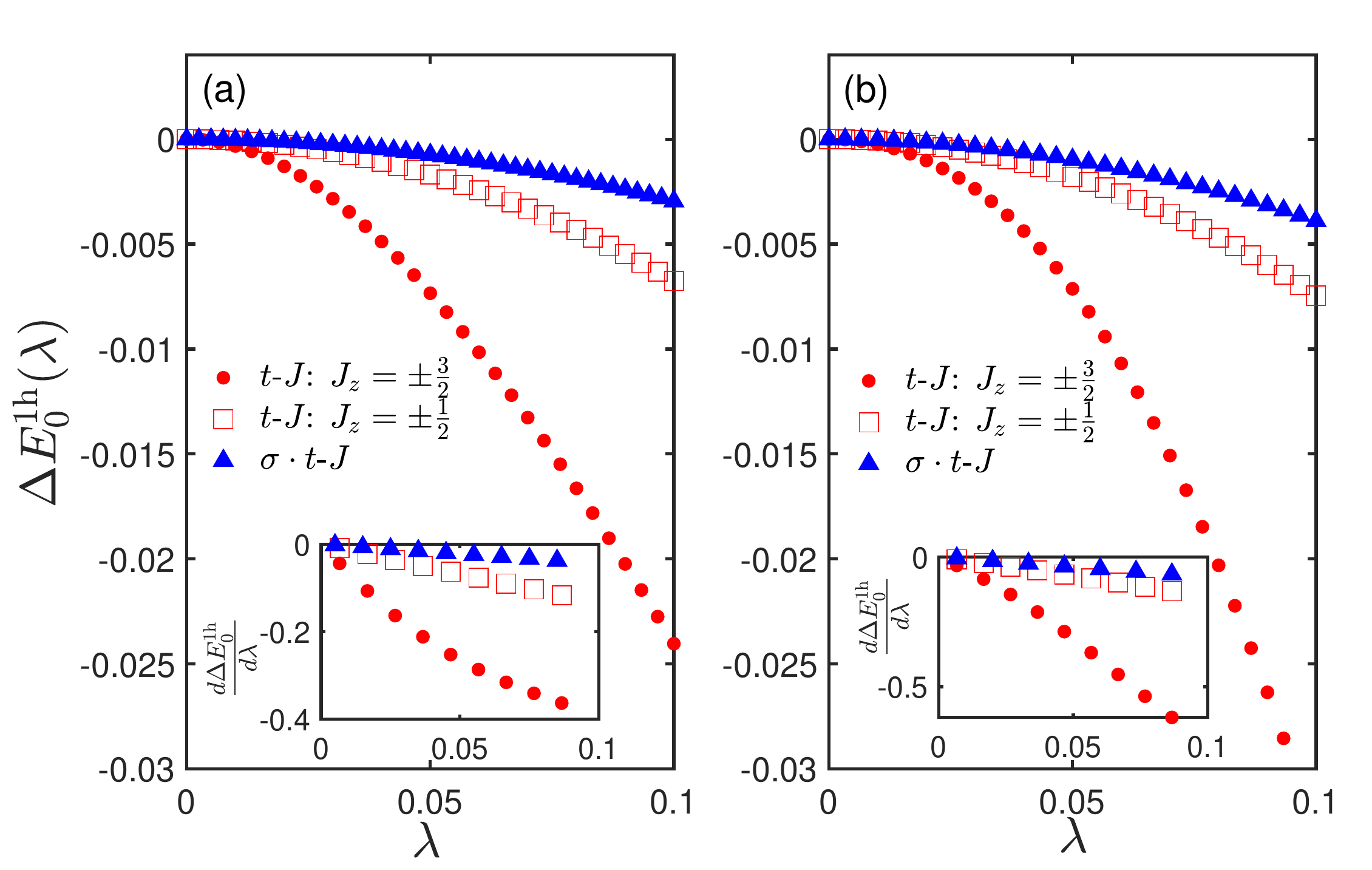} 
\caption{The energy shift $\Delta E_{0}^{\mathrm{1h}}$ due to the SOC for the degenerate ground states ($J_{z}=\pm\frac{3}{2}$)
and first excited states ($J_{z}=\pm\frac{1}{2}$) obtained by (a)
ED and (b) VMC calculations for the $t$-$J$ (red dot and red square)
and $\sigma\cdot t$-$J$ (blue triangle) models. Insets: the first derivation of $\Delta E_{0}^{\mathrm{1h}}$ 
over $\lambda$. Here $N=4\times4$ under OBC [cf. Fig.~\ref{Fig:tjPBCEnergyInten}(a) for ED under PBC]. }
\label{Fig:EnergyEDVMC} 
\end{figure}


In this work we shall examine some novel spin texture structure hidden
in the quasi-hole excitation of the doped Mott insulator described
by the $t$-$J$ model \cite{anderson1987theresonating,zhang1988effective}, whose Hamiltonian
is composed of the hopping $H_{t}$ and the superexchange $H_{J}$
terms as follows, 
\begin{equation}
\begin{aligned}H_{t} & =-t\sum_{\langle ij\rangle,\sigma}c_{i\sigma}^{\dagger}c_{j\sigma}^ {}+h.c.~,\\
H_{J} & =J\sum_{\langle ij\rangle} (\mathbf{S}_{i}\cdot\mathbf{S}_{j}-\frac{1}{4}n_{i}n_{j})~,
\end{aligned}
\label{t-J-ham}
\end{equation}
where $\langle ij\rangle$ denotes two nearest neighbor sites $i$
and $j$, and $c_{i\sigma}$ creates an electronic hole with the spin
index $\sigma=\uparrow$ or $\downarrow$ at site $i$. $\mathbf{S}_{i}$
and $n_{i}$ are the spin and electron number operators, respectively,
and the Hilbert space is always subject to the no double occupancy
constraint 
\begin{equation}
n_{i}\equiv\sum_{\sigma}c_{i\sigma}^{\dagger}c_{i\sigma}^ {}\leq 1~.
\end{equation}
At half-filling $n_{i}=1$, the system becomes the Mott insulator
with the spin degrees of freedom described by the Heisenberg Hamiltonian
$H_{J}$.
We fix the model parameters at $t/J=3$ and $J=1$ in the following.

The motion of the holes injected into such a system will strongly
interact with the quantum spin background to acquire a Berry-phase-like
emergent sign structure, i.e., the so-called \emph{phase string} \cite{weng1996phase,weng1997phase,wu2008sign}.
This singular effect is manifested by generating neutral spin current
\cite{zheng2018hidden,chen2019single} as the backflow associated
with the hopping of the doped holes. An important fact is that the
spin current is conserved, or the phase string created by the hopping
is ``irreparable'', which replaces the original fermion signs to
become the new statistical signs in the $t$-$J$ model \cite{wu2008sign}.
One may find a more detailed discussion on the phase string sign structure
in Appendix~\ref{App:signstru}.

On the other hand, the phase string sign structure in the doped $t$-$J$
model can be precisely eliminated by modifying $H_{t}$ into the following
form \cite{zhu2013strong,zhu2014nature}: 
\begin{equation}
H_{\sigma\cdot t}=-t\sum_{\left\langle ij\right\rangle }\sigma c_{i\sigma}^{\dag}c_{j\sigma}^ {}+h.c.~,
\label{sigmat}
\end{equation}
where a spin-dependent sign is inserted to the hopping process, while
the superexchange term $H_{J}$ remains unchanged. It can be shown
that the resulting $\sigma\cdot$$t$-$J$ model with $H_{\sigma\cdot t\text{-}J}\equiv H_{\sigma\cdot t}+H_{J}$
will no longer generate any spin currents as the phase string is precisely
``switched off'' [cf. Appendix \ref{App:signstru}]. Consequently
an injected hole will behave simply like a Landau quasiparticle with
no more internal structure associated with the backflow spin current. Thus, a comparative study of the $t$-$J$ and $\sigma\cdot $$t$-$J$ models can provide a unique understanding of the doped Mott physics.

\subsection{Spin-orbit coupling}

In this work, we shall introduce a Rashba SOC term as a perturbation, which is given as follows \cite{rashba2000theory}
\begin{equation}
H_{\text{R}}=\lambda\sum_{i}i(c_{i\uparrow}^{\dag}c_{i+\hat{y}\downarrow}^ {}+c_{i\downarrow}^{\dag}c_{i+\hat{y}\uparrow}^ {})-(c_{i\uparrow}^{\dag}c_{i+\hat{x}\downarrow}^ {}-c_{i\downarrow}^{\dag}c_{i+\hat{x}\uparrow}^ {})+h.c.~.
\label{RashbaH}
\end{equation}
Here the parameter $\lambda$ characterizes the Rashba SOC strength
(valued in units of $J=1$) and $\hat{x}\left(\hat{y}\right)$ denotes
the unit vector along the $x\left(y\right)$ direction. 
It is noted that a weak Rashba SOC in the cuprate may actually arise from the lack of the inversion symmetry in a double-layer system \cite{gotlib2018revealing}.  In the present approach, we shall neglect the interlayer $t$-$J$ and SOC coupling between the double layers. Namely one is still focused on the one layer problem with only retaining an effective weak intra-layer SOC given in Eq.~\eqref{RashbaH}, while the another layer experiences an opposite SOC \cite{gotlib2018revealing}.
As a perturbation
to the $t$-$J$ or $\sigma\cdot $$t$-$J$ model, $H_{\text{R}}$ can
further provide a locking between the spin and orbital angular momenta
in the new Hamiltonian $H_{t\text{-}J\text{-R}}=H_{t}+H_{J}+H_{\text{R}}$
or $H_{\sigma\cdot t\text{-}J\text{-R}}=H_{\sigma\cdot t}+H_{J}+H_{\text{R}}$.

Then the single hole problem previously explored by both ED and VMC
methods can be easily generalized to the present case, which will
reveal interesting properties of the doped hole as a new twisted
quasiparticle in a rather unique way.

\begin{figure}
\centering \includegraphics[scale=0.6]{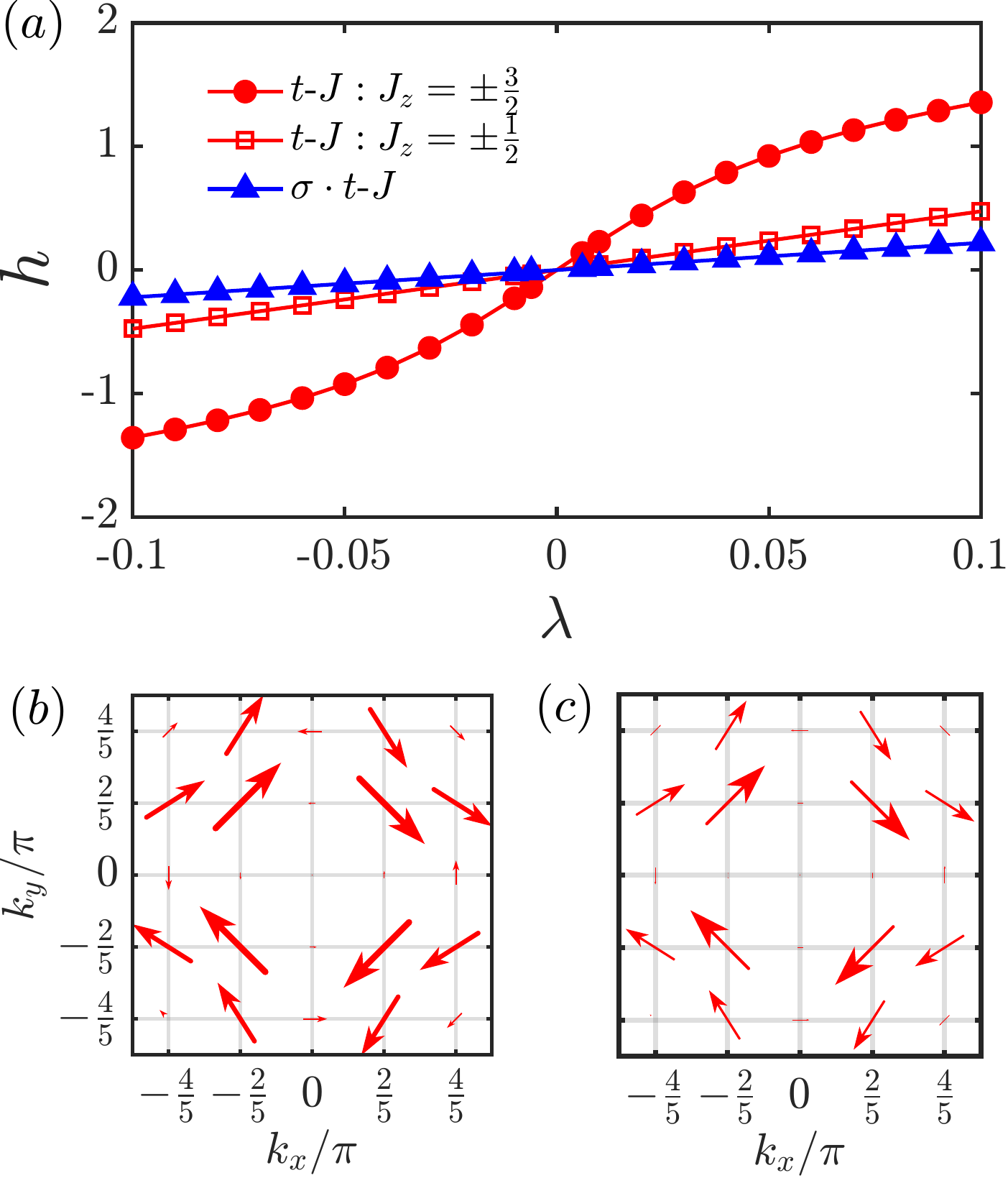} 
\caption{
The spin texture in momentum space. (a) The helicity $h$ defined in Eq.~\eqref{helicity} characterizes the overall strength of the spin texture; (b) and (c)
The corresponding momentum-space spin textures in the ground states ($J_{z}=\pm3/2$)
and the first excited states ($J_{z}=\pm1/2$) with the same chirality
$\chi=+1$ at a positive $\lambda=0.006$, calculated by ED ($N=4\times4$
under OBC). }
\label{Fig:helicity} 
\end{figure}

\begin{figure}[t]
\begin{centering}
\includegraphics[scale=0.6]{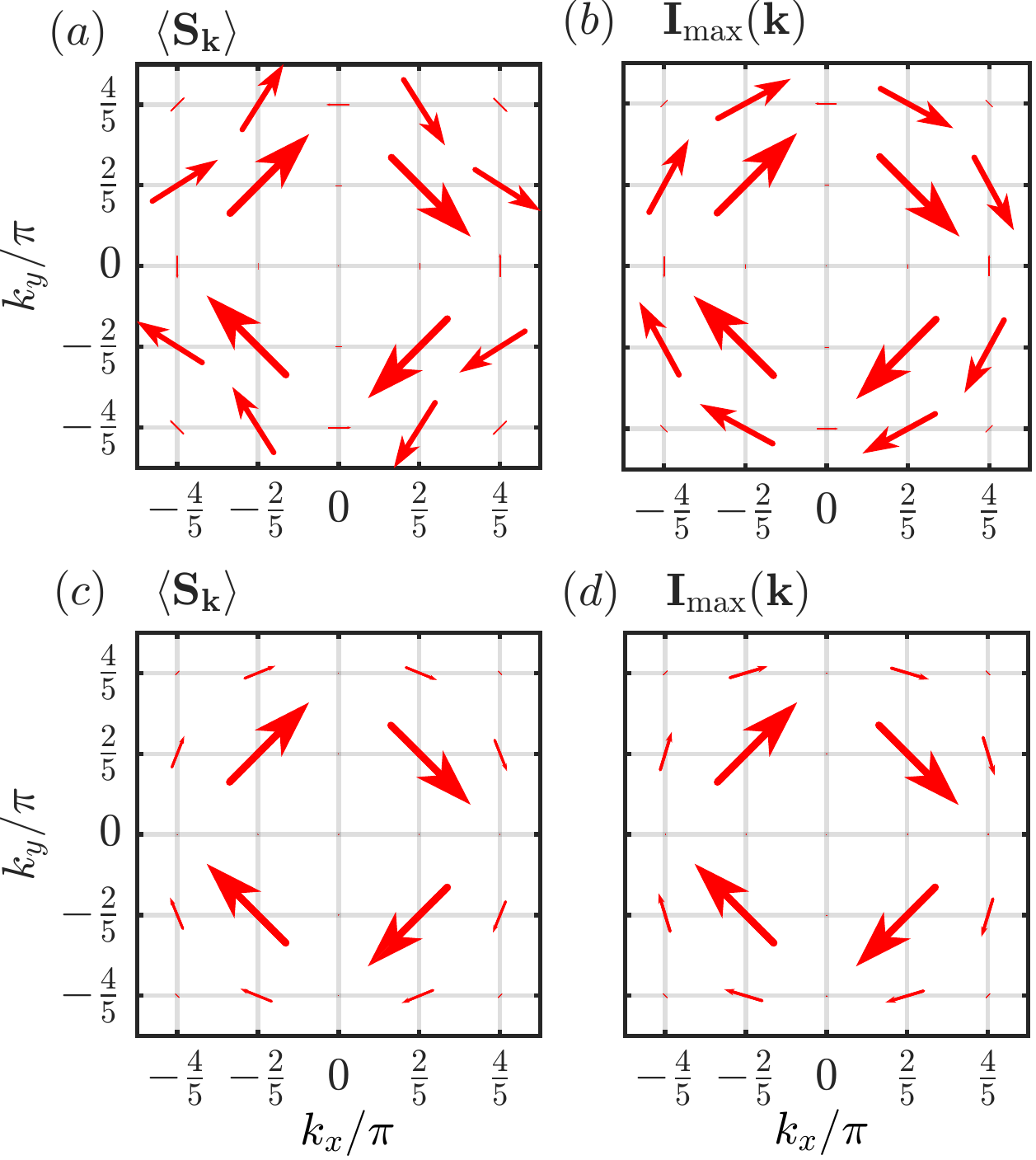} 
\par
\end{centering}
\caption{The spin texture in momentum space. (a) and (b): ED results of the
spin polarization $\langle\mathbf{S}_{\mathbf{k}}\rangle$ lying in
the $x$-$y$ plane in the ground state and the spin-polarized spectral
function $\mathbf{I}_{\mathrm{max}}(\mathbf{k})$, respectively; (c)
and (d): The corresponding VMC results of the variational state ansatz.
Here $\lambda=0.006$ in the $t\text{-}J\text{-R}$ model on a $4\times4$
lattice under OBC}
\label{Fig:OBCvortex} 
\end{figure}

\section{The Single Hole Study }
\label{sec:1holestudy}

As stated in the Introduction, the single-hole ground state of the
$t$-$J$ model has exhibited some highly nontrivial phenomenon as
shown by ED and DMRG \cite{zheng2018hidden}. Namely, the motion of
the hole in the $t$-$J$ model is always accompanied by a backflow
spin current due to the phase string effect, which then leads to a
continuous distribution of momentum for the hole as if there is an
incoherent many-body component in addition to a quasiparticle component
in a finite size sample. In the following, we shall examine the response
of such a two-component single-hole state to a weak Rashba SOC perturbation
by ED and VMC, respectively, which will then reveal the nature of
the internal structure of a doped hole as a unconventional many-body
composite in an explicit way.

\subsection{Exact diagonalization and variational ansatz in the absence of SOC
($\lambda=0$)}

To begin with, we note that for one hole injected into a half-filled
spin background in 2D, recent ED and DMRG studies in the pure $t$-$J$
model have clearly demonstrated \cite{zheng2018hidden} that the ground
state is characterized by an angular momentum $L_{z}=\pm1$ and $S^{z}=\pm1/2$
under OBC. In other words, the single hole eigenstates are totally
four-fold degenerate, which is not only true for the ground state
but also valid for the excited states as indicated by ED \cite{zheng2018hidden}.
It shows that the doped hole acquires an internal structure, which
is associated with a chiral spin currents. This is in contrast to
a Landau's quasiparticle where the many-body effect only leads to
a renormalization of its inertia by changing the effective mass without
affecting the translational symmetry and $U(1)$ charge nor inducing
an internal degree of freedom.

Such a novel structure associated with the doped hole can be captured
by a VMC wave function ansatz \cite{weng2011mott,weng2011superconducting,wang2015variational,chen2019single}. 
To understand the unique characterization of this wave function ansatz,
one may first recall a conventional Bloch-wave state given by 
\begin{equation}
|\Psi_{0}\rangle_{\mathrm{1h}}=\sum_{i}\varphi_{\mathrm{h}}^{0}\left(i\right){c}_{i\alpha}|\phi_{0}\rangle~,\label{landau}
\end{equation}
where the single-hole wave function $\varphi_{\mathrm{h}}^{0}$ is
a linear combination of the plane wave $\propto e^{\pm i{\bf k}\cdot{\bf r}_{i}}$
for a Wannier state ${c}_{i\alpha}|\phi_{0}\rangle$.
The latter with a single hole at site $i$ is created by annihilating
an electron from the translationally invariant half-filling ground
state $|\phi_{0}\rangle$, which may be best approximated by a \emph{bosonic}
resonant-valence bond (RVB) state \cite{liang1988some} 
with the spin-spin correlation length $\xi\rightarrow\infty$. If
the true single-hole ground state with a total momentum ${\bf k}$
can be adiabatically connected to Eq.~(\ref{landau}) with a finite
spectral weight $Z_{{\bf k}}$, we may call the hole state a Landau-like
quasiparticle.

However, instead of Eq.~(\ref{landau}), a novel single-hole state
wave function has been found \cite{chen2019single} in good agreement
with the above ED and DMRG results \cite{zheng2018hidden}. Mathematically
it is realized by replacing the bare hole creation operator $c$ in
Eq.~(\ref{landau}) with $\tilde{c}$: 
\begin{equation}
c_{i\sigma}\rightarrow\tilde{c}_{i\sigma}\equiv c_{i\sigma}e^{-i\hat{\Omega}_{i}}~~,\label{eqn:ctilde}
\end{equation}
which leads to a new single-hole ansatz state \cite{wang2015variational,chen2019single}
\begin{equation}
|\Psi\rangle_{\mathrm{1h}}=\sum_{i}\varphi_{\mathrm{h}}\left(i\right){c}_{i\alpha}e^{-i\hat{\Omega}_{i}}|\phi_{0}\rangle~.\label{VMC1h}
\end{equation}
Here the nontrivial phase shift operator takes the form 
\begin{equation}
\hat{\Omega}_{i}=\sum_{l}\operatorname{Im}\ln(z_{i}-z_{l})n_{l\downarrow}~,\label{phaseO}
\end{equation}
where $n_{l\downarrow}$ denotes the $\downarrow$-spin operator acting
on $|\phi_{0}\rangle$ with $z_{l}$ as the complex coordinate of
site $l$.
It encodes the mutual braiding between the hole
and the $\downarrow$-spins in the background spins, 
as illustrated in Figs.~\ref{Fig:schematicplot}, to give rise to
a nontrivial angular momentum $L_{z}=\pm1$ and a \emph{net} spin current \cite{chen2019single}
\begin{equation}
J_{ij}^s=i(S_i^+S_j^--S_i^-S_j^+)~.
\label{Jsr}
\end{equation}
The numerical ED and DMRG (up to $8\times8$) results \cite{zheng2018hidden}
for the $t$-$J$ model can be well reproduced by such a variational
ground state ansatz in Eq.~\eqref{VMC1h}, with the hole wave function
$\varphi_{\mathrm{h}}\left(i\right)$ being the variational parameter
in the VMC calculation \cite{chen2019single}. In Eq.~\eqref{eqn:ctilde},
a minus sign is chosen in $\hat{\Omega}_i$, which defines one possible chirality of the ground states.
One may also choose a positive sign, i.e. $e^{+i\hat \Omega_i}$, to accordingly give rise to a degenerate ground state with 
the opposite chirality. A detailed symmetry analysis of the wave function in Eq. (\ref{VMC1h}) is given in Appendix~\ref{App:Symmetry}. 

By contrast, the Landau-like wave function in Eq.~(\ref{landau})
is shown \cite{chen2019single} to be a good ground state for the
$\sigma\cdot$$t$-$J$ model, in which the phase-string sign structure
is switched off without creating a spin-current backflow during the
hopping process [cf. Fig.~\ref{Fig:schematicplot}(c)] even though
the spin $S_{z}=\pm1/2$ has a similar distribution around the hole
as in the $t$-$J$ case [cf. Fig.~\ref{Fig:schematicplot}(b)].
Obviously the two wave functions in Eq.~(\ref{landau}) and Eq.~\eqref{VMC1h}
cannot be adiabatically connected to each other, indicating the qualitative
distinction and non-perturbative nature of the single-hole state in
comparison with the conventional Landau paradigm. Crucially, the nonlocal
phase structure in Eq.~\eqref{eqn:ctilde} persists through the whole
spectrum of the single-hole eigenstates irrespective of the excited
energy. One may also continuously tune the spin-spin correlation length
$\xi$ in $|\phi_{0}\rangle$ variationally (see Appendix~\ref{APP:VMChf})
to simulate the case of an arbitrary finite $\xi$, which should be
self-consistently realized in the ground state at finite doping of
holes \cite{weng2011superconducting,weng2011mott}. The detailed structures
of the single-hole state as variationally obtained at $\xi\gg L$ and $\xi\sim 2.7$
are illustrated in Figs.~\ref{Fig:schematicplot}(b) and \ref{Fig:schematicplot}(c).

After the above brief outline of the single hole state in the pure case,
in the following, one can straightforwardly generalize the above calculations
to $\lambda\neq 0$ and study $H_{t\text{-}J\text{-R}}$ by the VMC
approach in comparison with the ED results. We will see that the exotic
non-Landau-like features of the doped hole will further show up.

\subsection{The single hole state under a weak Rashba spin-orbit coupling ($\lambda\protect\neq0$)}
\subsubsection{Exact diagonalization}
\label{ED}

Let us first start with the ED study under OBC, which is to be compared
with a VMC study based on an analytic ground state ansatz below. In
contrast to the ED under the periodic boundary condition (PBC) given
in Appendix~\ref{App:PBC}, the ground state degeneracy of the $t$-$J$
model (without SOC) under the OBC (with $C_{4}$ symmetry) has been
shown to be reduced to four-fold characterized by the orbit angular
momentum $L_{z}=\pm1$ and spin $S_{z}=\pm1/2$ \cite{chen2019single}. 
Once $\lambda$ is switched on in $H_{t\text{-}J\text{-R}}$, $S_{z}$
and the angular momentum $L_{z}$ are no longer conserved separately.
Instead, the total angular momentum $J_{z}=L_{z}+S_{z}$ will
remain as a good quantum number (cf. Appendix~\ref{App:AngMom}), with the ground states characterized
by $J_{z}=L_{z}+S_{z}=\pm3/2$ with two-fold remaining degeneracy by ED.
To show the overall strength of the SOC, the change of the ground
state energy as a function of $\lambda$ is shown in Fig.~\ref{Fig:EnergyEDVMC},
in which (a) is for ED and (b) is for VMC (see below). The energy
gain induced by SOC as well as the energy splitting between $J_{z}=\pm3/2$
and $J_{z}=\pm1/2$ is clearly shown here. By contrast, the SOC effect
for a quasiparticle in the $\sigma\cdot $$t$-$J$ model (blue triangles)
is much reduced. A similar result of the ED calculation under PBC
can be found in Appendix~\ref{App:PBC}.

To quantitatively characterize the spin texture induced by $\lambda\neq0$,
one may define a ``helicity'' as the projection of the averaged
spin $\langle\mathbf{S}_{\mathbf{k}}\rangle$ at the direction perpendicular
to the momentum $\mathbf{k}$ lying in the $x$-$y$ plane: 
\begin{equation}
h\equiv\sum_{\mathbf{k}}\frac{\langle\mathbf{S}_{\mathbf{k}}\rangle\cdot\mathbf{(}\mathbf{k}\times\hat{\mathbf{z}})}{|\mathbf{k}|}~, \label{helicity}
\end{equation}
where $\mathbf S_\mathbf k= \frac{1}{2}c_\mathbf{k}^\dagger \sigma c_\mathbf{k}$ with $c_\mathbf{k}=[c_{\mathbf k\uparrow},c_{\mathbf k\downarrow}]^T$.
It measures the strength of locking between the spin and orbit associated
with the single hole in the momentum space as shown in Fig.~\ref{Fig:helicity}(a),
with the sign $\chi=\mathrm{sign}(h)$ depicting the chirality of
the corresponding spin texture in Figs.~\ref{Fig:helicity}(b) and
(c). Here the helicity $h$ for $J_{z}=\pm3/2$ (the ground states)
shows a significantly larger value as compared to that of the first
excited state at $J_{z}=\pm1/2$ in Figs.~\ref{Fig:helicity}(b)
and (c), which both possess the same chirality $\chi$ at $\lambda>0$.
The sign of $\chi$ does not depend on the sign of $J_{z}$ but only
on the sign of $\lambda$ as shown in Fig.~\ref{Fig:helicity}(a).
It is easy to understand that the four spin textures in Figs.~\ref{Fig:helicity}(b)-(c)
and their counterparts at $\lambda<0$ just correspond to the eight
degenerate eigenstates of the $t$-$J$ model at $\lambda\rightarrow0$
in the thermodynamic limit or under PBC \cite{zheng2018hidden,chen2019single},
which are split into ground state and the first excitations in a finite
lattice under OBC, and further split at $\lambda\neq0$ with distinct
chiralities. On the other hand, the helicity $h$ for the $\sigma\cdot$$t$-$J$
model is indeed much reduced as shown in Fig.~\ref{Fig:helicity}(a).


\begin{figure}[t]
\centering \includegraphics[scale=0.5]{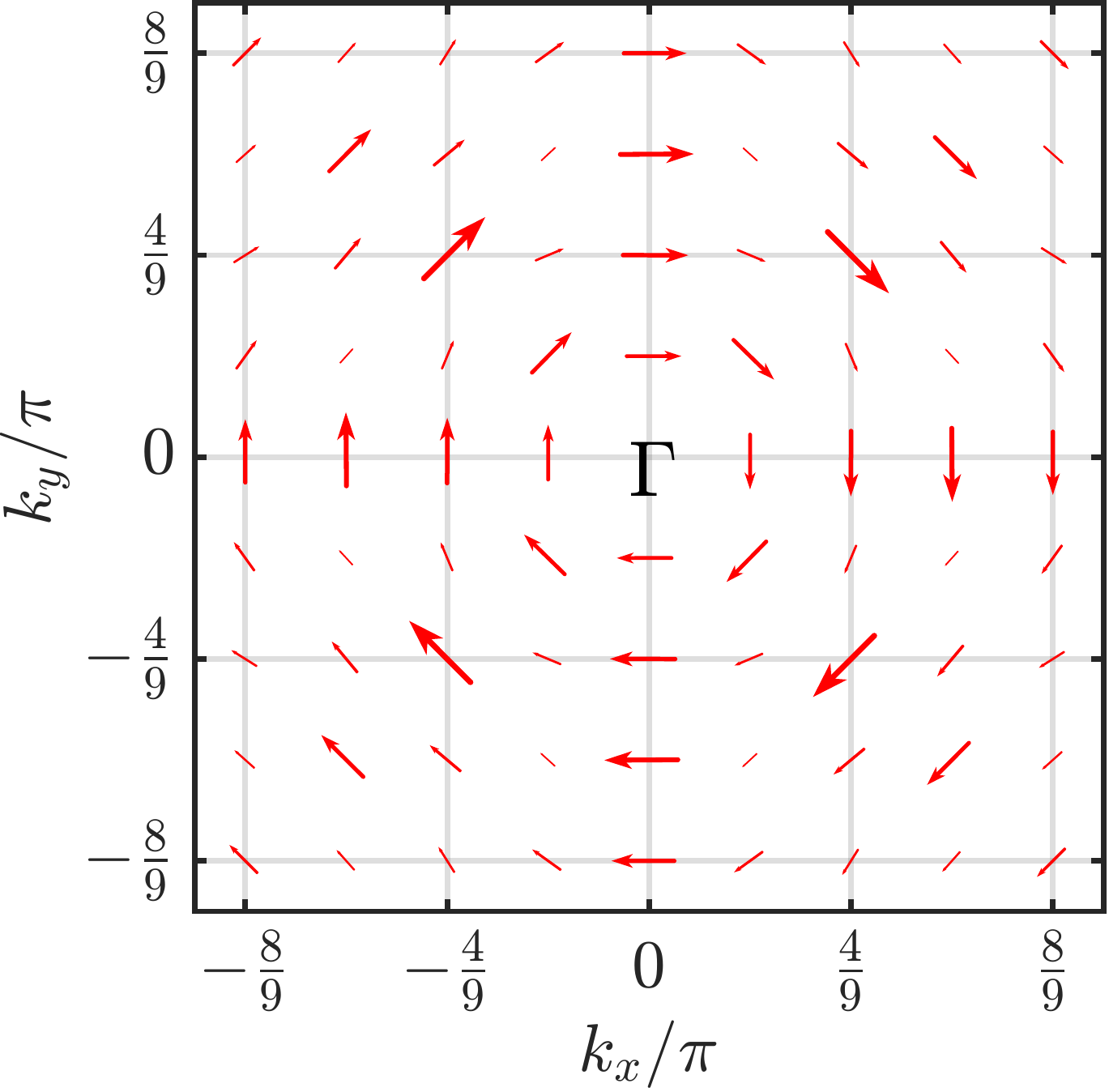}
\caption{Spin texture in momentum space as manifested in the spin-polarized spectral function ${\bf I}_{\mathrm{max}}({\bf k})$, which involves the excited states as
calculated by VMC in a lattice of $8\times8$ and $\lambda=0.01$.}
\label{Fig:8by8} 
\end{figure}

\subsubsection{Variational Monte Carlo calculation}

\label{Sec:VMC}

The above ED calculation shows that by introducing a weak Rashba SOC
to lift the separate $S_{z}$ and $L_{z}$ degeneracy, the eigenstate
degeneracy associated with the hole will be reduced to two with the
locking of $S_{z}$ and $L_{z}$ in the $t$-$J$ model. Specifically,
the total angular momentum $J_{z}\equiv L_{z}+S_{z}$, which
remains to be a good quantum number, will characterize the composite
hole $\tilde{c}$ as a \emph{pseduo-spin} in replacing $S_{z}=\pm1/2$.
Namely, the Rashba SOC will select a new quantum number $J_{z}$,
which leads to a chiral spin texture. A detailed symmetry analysis
is given in Appendix~\ref{App:AngMom}.

In the following, we shall employ a variational wave function to understand
the underlying physics involving the many-body response to a weak
SOC in the one-hole-doped Mott insulator. By generalizing the previous
four-fold degenerate wave functions for the pure $t$-$J$ model,
one may construct the following time-reversal invariant new ground
states in terms of the linear combinations of Eq.~(\ref{VMC1h}):
\begin{equation}
    |\Psi_{\mathrm{R}}\rangle_{\mathrm{1h}} =\!\!\!\sum_{i}[\varphi_{h}^{\downarrow}(i)e^{-i\hat{\Omega}_{i}}c_{i\downarrow}+\varphi_{h}^{\uparrow}(i)e^{- i\hat{\Omega}_{i}}c_{i\uparrow}]|\phi_{0}\rangle~,
    \label{wfansatz-so}
\end{equation}    
where $\varphi_{h}^{\uparrow}$ and $\varphi_{h}^{\downarrow}$ are
the variational parameters. One may also construct other 
types of ansatz by choosing a different factor $e^{\pm \Omega_i}$ in Eq.~(\ref{wfansatz-so}),
which variationally produces excited states. The VMC results are presented as follows 
[see Appendix~\ref{APP:VMCtJR} for the detail].

A spin texture around the hole has been clearly found to lie in the
$x$-$y$ plane in Fig.~\ref{Fig:spintexture}(a)
with $J_{z}=\pm3/2$ for the $t$-$J$ model (with $\lambda=0.01$).
Here the spin texture emerges out of the hidden spin current structure
of the non-Landau ``twisted'' hole in Fig.~\ref{Fig:schematicplot}(a)
via SOC. Its scale can be fitted by $\xi_{sh}\simeq\xi$
[cf. the inset of Fig.~\ref{Fig:schematicplot}(b)] with $\xi_{sh}$
being the spin-hole correlation length. In Fig.~\ref{Fig:spintexture},
a finite spin-spin correlation length $\xi\sim2.7$ in $|\phi_{0}\rangle$
has been used in the VMC. By comparison, one can examine the spin
texture in a Landau-like quasiparticle in the $\sigma\cdot$$t$-$J$
model, i.e., by turning off the phase string effect in the $t$-$J$
model. Without creating a spin current backflow, a doped hole will
propagate like a conventional Bloch wave in the $\sigma\cdot $$t$-$J$
model. At the same strength of $\lambda=0.01$, there is no more internal
spin texture structure around the hole [see the right panel of Fig.~\ref{Fig:spintexture}(b)].
Here the variational wave function of the $\sigma\cdot $$t$-$J$
model is given by 
\begin{equation}
|\Psi\rangle_{\sigma\cdot t\text{-}J\text{-}\mathrm{R}}=\sum_{i}[\varphi_{h}^{\downarrow}(i)c_{i\downarrow}+\varphi_{h}^{\uparrow}(i)c_{i\uparrow}]|\phi_{0}\rangle~,
\end{equation}
which is different from those of the $t$-$J$ model by simply switching
off the phase shift operator $\hat{\Omega}_{i}$ in Eq.~(\ref{wfansatz-so}).

The overall agreement of the energies between the ED and VMC calculations
are further illustrated in Fig.~\ref{Fig:EnergyEDVMC}, including
the energy split between $J_{z}=\pm3/2$ and $J_{z}=\pm1/2$ in the
$t$-$J$ model and the ground state energy shift in the $\sigma\cdot$$t$-$J$
model. One may see the contributions of the SOC to the variational
energy shifts in Fig.~\ref{Fig:EnergyEDVMC}(b) as a function of
$\lambda$, which are in excellent agreement with the ED results [Fig.~\ref{Fig:EnergyEDVMC}(a)].
For the $\sigma\cdot $$t$-$J$ model, in which a Landau's quasiparticle
picture can be recovered at $\lambda=0$ \cite{chen2019single}, one
finds a drastically weakened SOC effect at the same $\lambda$ as
compared to the $t$-$J$ model. As pointed out above, the latter
only differs from the former by the phase string sign structure as
represented by the backflow spin current accompanying the hopping.
Similar contrast has been also clearly shown by ED under PBC as presented
in Appendix~\ref{App:PBC}. Therefore, the phase string effect indeed
significantly amplifies the weak Rashba SOC effect in the $t$-$J$ model.

Furthermore, given the spatial spin texture around the hole in Fig.~\ref{Fig:spintexture}(a)
, a corresponding spin texture described by the average
spin $\langle\mathbf{S}_{\mathbf{k}}\rangle$ in momentum space
is expected by the rotational symmetry, which has been already illustrated in, say,
Fig.~\ref{Fig:helicity}(b) at $J_{z}=\pm3/2$. In Fig.~\ref{Fig:OBCvortex},
the spin-momentum-locking spin textures obtained by ED and VMC are
presented together for comparison, in which $\langle\mathbf{S}_{\mathbf{k}}\rangle$
in the ground state are shown in (a) and (c) for a $4\times4$ lattice.

In Figs.~\ref{Fig:OBCvortex}(b) and (d), we further present the
ED and VMC results for the spin-polarized spectral function ${\bf I}_{\mathrm{max}}({\bf k})$,
which is defined in Eq.~(\ref{IntenVect}) of Appendix~\ref{Sec:Spectfun}
at each given momentum $\mathbf{k}$ with the maximal magnitude
of the spectral function via properly adjusting the excited energy.
In other words, ${\bf I}_{\mathrm{max}}({\bf k})$ can effectively
detect the maximal spin texture in momentum space involving the excited
states, which is experimentally relevant. In Fig.~\ref{Fig:8by8}, the spin texture in momentum space
is calculated by VMC in a larger ($8\times8$) lattice such that more
momentum points are available, where the spin vortex structure is
clearly illustrated in the full Brillouin zone, or in other words,
persists over the high energy towards the $\Gamma$ point.

Finally, we emphasize that it is the intrinsic phase string that causes the hole dressed by a surrounding spin current of the $S=1/2$ partner and the weak Rashba SOC only lifts the degeneracy to induce a spin texture by breaking the spin rotational symmetry.  If one starts with the large-U Hubbard model, the high order corrections to the $t$-$J$ model may be at the same order as $\lambda$, but they do not break the spin rotational symmetry like the SOC does and it has been shown \cite{Zhanglong2014} that the phase string effect still present. The present conclusion should thus be expected to remain unchanged.

\section{Discussion: Finite Doping}
\label{sec:finitedoping}
For the single hole case studied in the previous section, the spin-1/2
partner forms a spin current vortex surrounding the hole as illustrated
in Fig.~\ref{Fig:schematicplot}, which is then ``frozen''
into a spin texture lying in the $x$-$y$ plane via a weak SOC as
shown in Fig.~\ref{Fig:spintexture}(a). The
excellent agreement between the ED and VMC methods clearly indicates
that the single-hole ansatz wave function well captures the \emph{intrinsic}
structure of the doped hole in a finite size lattice that cannot be
adiabatically connected to a Landau-like quasiparticle, not only in
the absence of SOC but also in the presence of a weak Rashba SOC.
\begin{figure}[t]
\centering \includegraphics[scale=0.8]{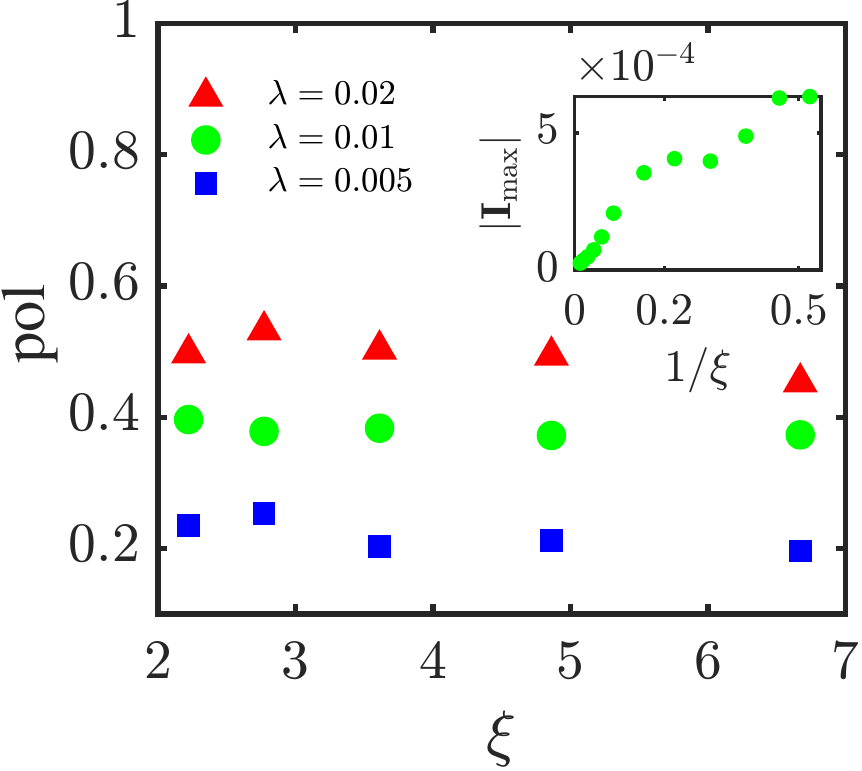} 
\caption{The relative strength $\mathrm{pol}(\omega_{c},\mathbf{k})$ of the spectral function with spin polarized perpendicular to the momentum [defined in Eq.~\eqref{pol}] at a given $\mathbf{k}=(\pi/2,\pi/2)$. Here the spin correlation length $\xi$ is taken in a finite range at different $\lambda$'s (see the text). Inset: The full 
magnitude of the spectral function $|{\bf I}_{\mathrm{max}}({\bf k})|$ as a function of $\xi$.
The sample size is $16\times16$. 
}
\label{Fig:fintedoping} 
\end{figure}


Recently, the VMC approach has been further generalized to the two-hole states in the absence of SOC \cite{chen2018twohole,zhao2021two}. By forming a tightly bound pair, the chiral spin current gets cancelled and the ground state becomes non-degenerate with angular momentum $L_z=2$ and a d-wave pairing symmetry [cf. Ref.~\onlinecite{zhao2021two}], which is also consistent with the ED and DMRG result \cite{zheng2018hidden}. According to the above discussion, in the absence of the chiral spin current, a weak SOC usually should not be able to induce a spin texture in a pairing state of holes in contrast to a single hole case. Therefore, at finite doping, the spin texture associated with a doped hole is expected to appear only when the unpaired holes are present in an excited state.

Two questions remain to be answered. One concerns the SOC effect in
the thermodynamic limit, e.g., in the AF long-range ordered state
with $N=L^{2}\rightarrow\infty$ and $\xi\rightarrow\infty$, where
ED certainly does not apply. The VMC calculation indicates that the
SOC effect should get diminished continuously as the spin current
associated with the spin-1/2 spreads over in space with the increase
of $\xi_{sh}\simeq\xi$. However, in a true AF long-range ordered state, the gapless spin excitations
from the spin background can be further excited to form an even stronger spin currents around the hole. In such a limit, the strong spin fluctuations may even cause the self-localization of the doped hole \cite{chen2019single}, leading to a very singular effect in response to an SOC, which is beyond the present VMC approach at a finite size.  
On the other hand, the spin texture as the internal structure of the
doped hole discussed in this work is expected to be robust if $\xi$ becomes finite in the
spin background with $L\gg\xi$. What we have explored here is actually about how a hole moving in a spin background of finite $\xi$. For example, the ED results of a single hole on a $4\times 4$ lattice may be roughly understood as a finite doping $\sim 1/16$ system. As a matter of fact, the variational single hole state in Eq.~(\ref{VMC1h}) may be created in a spin
background $|\phi_{0}\rangle$ whose spin-spin correlation length
$\xi$ can be artificially tuned (cf. Appendix \ref{APP:VMChf}). For instance, to
better illustrate the spin texture around the hole in Fig.~\ref{Fig:spintexture}(a),
a variational $|\phi_{0}\rangle$ with a finite $\xi\sim2.7$ has
been used. Since $\xi_{sh}\simeq\xi\ll L$ in this case, the spin
texture as the integral part of the hole composite is clearly shown
in Fig.~\ref{Fig:spintexture}(a). In other words,
if a finite doping mainly changes the spin background into a short-range
AF state (or spin liquid) with the doped hole still described by $\tilde{c}_{i\sigma}$
in Eq.~(\ref{eqn:ctilde}), the above account of the Rashba SOC effect
should be valid in the leading order of approximation.

So the second question will be whether the present study of the single-hole
behavior under the Rashba SOC can be meaningfully generalized to the
quasiparticle excitation at finite doping, where a short-range AF
state with finite $\xi$ will set in. Namely, a single-particle excitation
above the ground state $|\Psi_{\mathrm{G}}\rangle$ (which may be
either a superconducting or pseudogap state at finite doping, see
below) may be constructed by 
\begin{equation}
|\psi\rangle_{\mathrm{qp}}=\sum_{i}\varphi_{\mathrm{h}}(i)\tilde{c}_{i\sigma}|\Psi_{\mathrm{G}}\rangle~,\label{qp}
\end{equation}
similar to the one-hole state in Eq.~(\ref{VMC1h}) with replacing
the ``vacuum'' state $\vert\phi_{0}\rangle$ by $\vert\Psi_{\mathrm{G}}\rangle$.
We shall delay the discussion of such a second question for a moment.
Let us first assume that Eq.~(\ref{qp}) is indeed valid, in which
$|\Psi_{\mathrm{G}}\rangle$ may be still treated as a short-ranged
AF state $|\phi_{0}\rangle$ with the spin correlation length $\xi$
tunable. Then one may further make an estimation of the strength of
the SOC effect on the quasiparticle excitation as follows. 

The spin-polarized spectral function corresponding to the quasiparticle
excitation in Eq.~(\ref{qp}) will be then approximately treated
similar to that of the single hole case under the SOC (cf.  Appendix~\ref{Sec:Spectfun}). The spin texture
profile, determined by VMC in the momentum space based on ${\mathbf{I}}_{\mathrm{max}}(\mathbf{k})$,
is similar to Fig.~\ref{Fig:8by8} with the magnitude depending on
$\xi$ and $\lambda$. In Fig.~\ref{Fig:fintedoping}, a relative
strength of the spin texture at $\mathbf{k}=(\pi/2,\pi/2)$ is shown
as a function of $\xi$ at various $\lambda$'s, which is defined by \cite{gotlib2018revealing}
\begin{equation}
\mathrm{pol}(\omega_{c},\mathbf{k})\equiv\frac{\mathcal{A}_{\sigma_{\perp}}(\omega_c,\mathbf k)-\mathcal{A}_{\bar{\sigma}_{\perp}}(\omega_c,\mathbf k)}{\mathcal{A}_{\sigma_{\perp}}(\omega_c,\mathbf k)+\mathcal{A}_{\bar{\sigma}_{\perp}}(\omega_c,\mathbf k)}~,\label{pol}
\end{equation}
in terms of the spin-polarized ARPES spectral function $\mathcal{A}_{\sigma}(\omega,\mathbf{k})$ at energy $\omega$ with spin index $\sigma$ [defined in Eq.~(\ref{SpectrumFunc}) in Appendix~\ref{Sec:Spectfun})]. Note
that here $\sigma_{\perp}$ denotes the spin polarization perpendicular
to $\mathbf{k}$ ($\bar{\sigma}_{\perp}=-\sigma_{\perp}$) and $\omega_{c}$
the frequency at which the intensity reaches its maximum. As Fig.~\ref{Fig:fintedoping} shows, $\mathrm{pol}(\omega_{c},\mathbf{k})$ is roughly flat in a range of finite $\xi$ with the strength monotonically dependent on $\lambda$, which is comparable to the experimental measurement in the cuprate \cite{gotlib2018revealing}. Note that $\mathrm{pol}$ here only measures the relative spin polarization perpendicular
to $\mathbf{k}$. In the inset of Fig.~\ref{Fig:fintedoping}, the total strength of the spectral function
$|{\mathbf{I}}_{\mathrm{max}}|$ (cf. Appendix~\ref{Sec:Spectfun}) at the same $\mathbf{k}=(\pi/2,\pi/2)$
is also shown as a function of $\xi$ as $\xi\rightarrow\infty$, which eventually approaches zero at the AF long-ranged
order limit. Physically it may be understood as the spin current carried by the spin-$1/2$ partner of the hole
composite becomes less and less concentrated around the hole to result
in the reducing SOC effect at large $\xi$ and $L$.

Now let us come back to address the above second question about the
single-particle excitation at finite doping. 
We point out that the single-hole ground state ansatz [Eq.~(\ref{VMC1h})]
can be actually regarded as the few hole special cases \cite{chen2019single,wang2015variational,chen2018twohole,zhao2021two}
of a general ground state ansatz previously proposed to describe the
SC and pseudogap phase at finite doping \cite{weng2011superconducting,weng2011mott}:
\begin{equation}
|\Psi_{\mathrm{G}}\rangle=\hat{{\cal D}}^{\frac{N^{h}}{2}}|\mathrm{RVB}\rangle~,\label{LPP}
\end{equation}
in which the AF vacuum state $|\phi_{0}\rangle$ is self-consistently
replaced by $|\mathrm{RVB}\rangle$, which is AF short-range-ordered
at a finite doping. Here the $N^{h}$ holes are paired up and created
by 
\begin{equation}
\hat{{\cal D}}=\sum_{ij}g_{ij}\tilde{c}_{i\uparrow}\tilde{c}_{j\downarrow}~,\label{D2}
\end{equation}
where $\tilde{c}$ defined by Eq.~(\ref{eqn:ctilde}) describes the
doped holes as twisted (non-Landau) quasiparticles. Here $|\Psi_{\mathrm{G}}\rangle$
is intrinsically superconducting at $T=0$ in a finite concentration
of holes. The excitations of free unpaired spinons in $|\mathrm{RVB}\rangle$
can eventually destroy the SC phase coherence, resulting in the so-called
lower pseudogap phase (LPP) or spontaneous vortex phase, where the
pairing in terms of the twisted quasiparticles, i.e., $\langle\hat{{\cal D}}\rangle\neq0$
is still present, but the true Cooper pairing order parameter vanishes
\cite{weng2011superconducting,weng2011mott}.

Naturally, an elementary quasiparticle excitation can be constructed,
based on $|\Psi_{\mathrm{G}}\rangle$, as composed of the unpaired
twisted quasiparticle given in Eq.~(\ref{qp}). In particular,
since the twisted quasiparticles are paired up via $\hat{{\cal D}}$
in $|\Psi_{\mathrm{G}}\rangle$, the latter may be reasonably considered
to be a ``vacuum" as far as $\tilde{c}$ is concerned in Eq.~(\ref{qp}),
which thus resembles a $|\phi_{0}\rangle$ just like the single-hole
state in Eq.~(\ref{VMC1h}) except for a finite $\xi$. Furthermore,
turning on the Rashba SOC in $H_{t\text{-}J\text{-R}}$ with $\lambda\neq0$,
there will be an additional important effect occurring in the SC/LPP
at finite doping, which can further enhance the spin texture structure
found in the single hole case, besides the reducing $\xi$ discussed
above, as to be outlined below.

Note that in the above SC/LPP ansatz states, the pairing between the
twisted quasi-holes of $\tilde{c}_{i\sigma}$ with $\sigma=\uparrow,\downarrow$
will be modified by the SOC with the degenerate ground states characterized
by a time-reversal pair of new good quantum numbers: $J_{z}=3/2$
and $J_{z}=-3/2$, i.e., 
\begin{equation}
\hat{{\cal D}}\rightarrow\sum_{ij}{g}(i,j)\tilde{c}_{i,J_{z}=3/2}^ {}\tilde{c}_{j,J_{z}=-3/2}^ {}~.\label{eqn:D1}
\end{equation}
As we have seen, the twisted hole states with both $J_{z}=\pm3/2$
have the same chirality of the spin texture. 
in Eq.~(\ref{eqn:D1}), the low-lying excitations will be mainly
dominated by those unpaired quasiparticles of $J_{z}=\pm3/2$ with
the same spin texture, where the other excited states with opposite
chirality in the single hole problem should no longer be stable to
contribute to the spin-polarized ${\mathbf{I}}_{\mathrm{max}}$ shown
in Fig.~\ref{Fig:8by8}. Thus, ${\mathbf{I}}_{\mathrm{max}}$ is
expected to be further enhanced due to the fact that the twisted
quasiparticles of $J_{z}=\pm3/2$ are uniquely singled out via the
pairing condensation in $|\Psi_{\mathrm{G}}\rangle$ to exclude all those
excitations of opposite spin textures split by the SOC.

\begin{figure}[t]
\centering \includegraphics[scale=0.8]{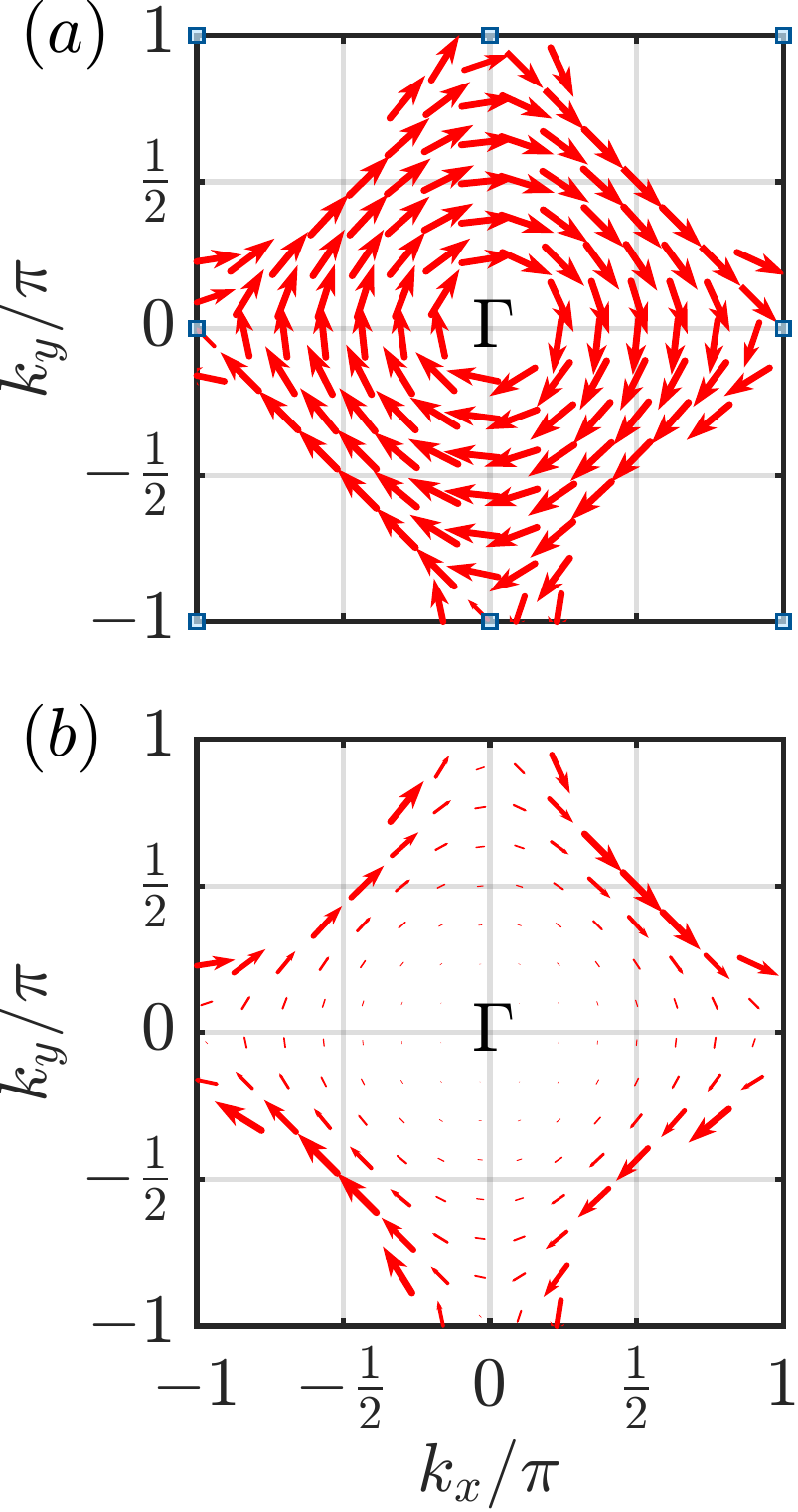} 
\caption{Spin texture configuration due to a weak Rashba SOC as determined by the spin-polarized spectral function ${\mathbf{I}}_{\mathrm{max}}(\mathbf{k})$ for (a) a free
Fermi gas, and (b) a Fermi liquid. The spin textures are fragile in
the Fermi liquid once the scattering between Laudau's quasiparticle
is taken into consideration (cf. Appendix~\ref{App:FL} for the detail). }
\label{Fig:SpinTextureFL} 
\end{figure}


\section{Concluding Remarks}
\label{sec:conclusion}
As the precursor of a non-Fermi liquid, it has been found recently
\cite{chen2019single,zhao2021two} that a hole injected into
a quantum AF spin background will behave like a non-Landau 
quasiparticle. It is a hole composite in which the neighboring spin background will get ``twisted'' by the 
phase string sign structure to facilitate the hopping of the hole. In the single-hole ansatz state (\ref{VMC1h}), such a many-body effect is well captured by the nonlocal phase shift factor $e^{-i\hat{\Omega}_i}$. In particular, such a novel structure can become explicitly manifested if an extra $S_z=\pm 1/2$ (introduced by the hole doped into the spin singlet background) is nearby in a finite system, which will form a spin current swirling around the hole like in an atom with a nontrivial angular momentum $L_{z}=\pm1$. 

Then we have shown in this work that by introducing a weak Rashba SOC
to lift the spin-1/2 degeneracy associated with the doped hole, the spin current will be turned into a semiclassical chiral spin texture lying in the $x$-$y$ plane in both real and momentum space, which is characterized
by a new quantum number $J_{z}=L_{z}+S_{z}$. It is important to emphasize that a weak SOC does not change the essential internal structure of the hole composite, but mainly affects the ``soft'' spin current contributed by the degenerate $S_z=\pm 1/2$, which are locked with $L_{z}$ due to breaking the spin rotational symmetry by SOC. In this sense, the present VMC approach is perturbative-like based on the \emph{good} (variational) solutions in the pure $t$-$J$ and $\sigma\cdot t$-$J$ models. In particular, the semiclassical spin texture formed by the spin-1/2 around the hole in Fig.~\ref{Fig:spintexture}(a) is merely a ``fingerprint'' that explicitly reveals the underlying quantum many-body hole-spin entanglement, which already exists in the absence of SOC. 
A recent proposal of the so-called rotational ARPES spectra \cite{PhysRevLett127197004} may be also used to probe such an internal spin structure associated with the doped hole.

By using ED and VMC methods, such a spin texture pattern characterized by the new quantum number $J_{z}$ has been
confirmed by both exact numerics and the variational wave function. The excellent agreement
between the ED and VMC results shows that the hole composite for the $t$-$J$ model has properly captured the essence of a doped Mott insulator in response to a perturbation of the SOC. On the other hand, by precisely turning off the phase string in the $\sigma\cdot t$-$J$ model, both ED and VMC calculations also demonstrate the disappearance of the
novelty of the non-Landau quasiparticle, with a substantially weakened spin texture entirely induced by the Rashba SOC for
a Landau-like quasiparticle.

Furthermore, we have argued that such a chiral spin texture in the
$t$-$J$ model may become directly observable at finite doping
in the superconducting/pseudogap regime, where the majority of the
doped holes form a paired condensate, which effectively selects the
hole composites with opposite quantum numbers $J_{z}=\pm3/2$ in a weak Rashba
interaction with the same chiral spin texture. In particular,
by making the spin AF background short-ranged, the hole composite
becomes ``atom-like'' as a truly bound entity such that the spin texture
around the hole is both amplified by the Rashba effect and independent
of the lattice size. Consequently, the low-lying quasiparticles excited
from the paired condensate will be predominately characterized
by the same spin texture structure in momentum space, which may be well observed by the spin-polarized
ARPES in momentum space, extending deep inside the Brillouin zone
away from the Fermi energy. Of course, we caution that the present discussion of the finite doping phase is mostly based on a generalization of the single hole problem.
A further microscopic study at the same level as dealing with the emergent Bogoliubove/Landau quasiparticle with Fermi arc \cite {zhang2020arpes} will be desirable to describe the spin texture structure under a weak SOC.

It is important to note that even though a similar spin texture
in momentum space (but not in the real space) commensurate with the
Rashba SOC can be simply found in a noninteracting
band structure model as shown in Fig.~\ref{Fig:SpinTextureFL}(a)
(cf. Appendix~\ref{App:FL}), the spin texture is actually well protected
only within a very narrow energy regime near the Fermi surface decided
by the SOC strength $\lambda$. As illustrated in Fig.~\ref{Fig:SpinTextureFL}(b),
the spin texture at momenta away from the Fermi surface becomes much
fragile in a Fermi liquid theory due to a general scattering effect
on the quasiparticles once the excitation energy is higher than a
small scale set by the SOC splitting energy (cf. Appendix~\ref{App:FL}).

We comment on a recent experimental result of the spin-polarized
ARPES experiment on the high-$T_{c}$ cuprate of the Bi-2212 compound
\cite{gotlib2018revealing} in the pseudogap regime with a $J\sim120$ meV. The dependence of the in-plane spin polarization
on the quasiparticle momentum observed in the experiment seems to
be compatible with a weak Rashba SOC strength as studied in this work,
which presumably comes from the double interlayer structure \cite{gotlib2018revealing}.
It is an important observation that the spin polarization extends
deep towards the $\Gamma$ point of the Brillouin zone, which is hard
to understand from a conventional Fermi liquid theory. 
As discussed
above, the spin texture found in the spin polarization measurement
should be restricted by a characteristic energy scale due to the SOC
of a single quasiparticle, which is rather small (e.g., $\lambda \sim 0.01 J \sim 1.2$ meV) and determines a very
narrow regime in the momentum space near the Fermi energy. 
By contrast, the spin texture of the new quasiparticle in a doped Mott
insulator is more intrinsically associated with the internal structure
of the doped hole, which is present even for the high energy excitations. An entirely different mechanism due to quantum
many-body correlations is at work here.

Finally, along with the SOC that breaks the spin rotation, we point out another important effect involving the Zeeman field, which also breaks the spin rotational symmetry. Such a Zeeman effect acts on the whole neutral spin background, in contrast to SOC mainly coupling a hole with its surrounding spins. Although no spin texture is expected, the Zeeman effect that polarizes spins can exert strong influence on the charge degree of freedom via an enhanced spin current due to the enlarged net spin $S_z$. Such an important effect due to phase string is currently under study elsewhere.

\begin{acknowledgments}
We acknowledge stimulating discussion with Jingyu Zhao. This work is supported by MOST of China under Grant No. 2017YFA0302902.
\\
 
\end{acknowledgments}

\newpage{}

\appendix

\section{Sign structures in the $t$-$J$ and $\sigma\cdot $$t$-$J$ models
\label{App:signstru}}

To be self-contained, in the following we briefly outline the phase-string
sign structure in the $t$-$J$ model and the absence of it in the
$\sigma\cdot $$t$-$J$ model \cite{weng1996phase,weng1997phase,wu2008sign}.

In the $t$-$J$ model, the Fermi sign structure vanishes at half-filling
due to the no-double-occupancy constraint or the ``Mottness''. Upon
doping, a reduced Fermi sign structure starts to recover with the
increase of doping concentration of holes. Specifically, the partition
function for the $t$-$J$ model on a square lattice at an arbitrary
doping can be expressed by 
\begin{equation}
Z_{t\text{-}J}=\sum_{c}\tau_{c}\mathcal{Z}[c]~,
\label{partitionZtj}
\end{equation}
where the summation is over $c$, denoting the set of closed loops
of the spatial trajectories of all spins and holes. Accompanying a non-negative
weight $\mathcal{Z}[c]\geq0$, each path $c$ is associated with a
sign factor $\tau_{c}$ given by 
\begin{equation}
\tau_{c}\equiv\left(-1\right)^{N_{h}^{\downarrow}\left[c\right]}\times\left(-1\right)^{N_{h}^{h}\left[c\right]}~.\label{tauc}
\end{equation}
Here the Berry's phase-like $\tau_{c}$ replaces the usual Fermion
signs in a partition function of a fermion system due to an explicit
restriction of the Mottness. In $\tau_{c}$, $\left(-1\right)^{N_{h}^{h}\left[c\right]}$
resembles a conventional Fermi sign structure now only associated
with the doped holes, while $\left(-1\right)^{N_{h}^{\downarrow}[c]}$
depends on the total parity of exchanges between the holes and $\downarrow$-spins
for the closed path set $c$, which is known as the phase-string sign
structure.

By contrast, in the $\sigma\cdot$$t$-$J$ model, the phase-string
sign structure can be simply \emph{switched off} such that the partition
function is reduced to 
\begin{equation}
Z_{\sigma\cdot t\text{-}J}=\sum_{c}\left(-1\right)^{N_{h}^{h}\left[c\right]}\mathcal{Z}\left[c\right]~,
\end{equation}
where the non-negative weight $\mathcal{Z}\left[c\right]$ remains
the \emph{same} as in the $t$-$J$ model, with the same Fermi sign
structure of the holes: $\left(-1\right)^{N_{h}^{h}\left[c\right]}$.
At half-filling, the $t$-$J$ and $\sigma\cdot$$t$-$J$ models
both reduce to the Heisenberg model.

Therefore, the distinction between the $t$-$J$ and $\sigma\cdot$$t$-$J$
models only emerges upon doping, which is solely distinguished by
the phase-string sign structure $\left(-1\right)^{N_{h}^{\downarrow}\left[c\right]}$
in Eq.~(\ref{tauc}). As it turns out, such a phase string is very
singular, which prevents a perturbative treatment of the $t$-$J$
model, even in the one-hole-doped case. In the next appendix, we shall
discuss a duality transformation and an emergent symmetry in the one-hole
ground state, which are entirely due to the phase string effect in
a non-perturbative way.

\section{One-hole wave function ansatz and emergent symmetry in the $t$-$J$ model }
\label{App:Symmetry}

The Mott physics in the pure $t$-$J$ model is crucially associated
with an emergent phase string sign structure as outlined above, which
will result in a non-perturbative duality transformation and a new
type of ground state ansatz \cite{wang2015variational,chen2019single}.
In the following, we concentrate on such a one-hole problem from the
perspective of symmetry.

To account for the phase string effect in the one hole case, a many-body
phase shift factor $e^{\pm i\hat{\Omega}_{i}}$ produced by the doped
hole has to be introduced \cite{wang2015variational,chen2019single}
in Eq.~(\ref{eqn:ctilde}), with $\hat{\Omega}_{i}$ defined in Eq.~(\ref{phaseO}).
The four-fold degenerate one-hole ground states can be constructed
with $|\Psi_{\mathrm{G}}\rangle_{1,2}$ by removing a $\downarrow$-spin
and $|\Psi_{\mathrm{G}}\rangle_{3,4}$ by removing an $\uparrow$-spin
from the half-filling ground state $|\phi_{0}\rangle$ as follows:
\begin{align}
|\Psi\rangle_{1,2} & =  \sum_{i}\varphi_{1,2}\left(i\right)e^{\mp i\hat{\Omega}_{i}}c_{i\downarrow}|\phi_{0}\rangle\text{ },\label{gsansatz1}\\
|\Psi\rangle_{3,4} & = \sum_{i}\varphi_{3,4}\left(i\right)e^{\mp i\hat{\Omega}_{i}}c_{i\uparrow}|\phi_{0}\rangle\text{ }.\label{gsansatz2}
\end{align}
Here the variational parameters $\varphi_{1,2,3,4}\left(i\right)$
only depend on the hole's position. The variational ansatzes in Eqs.~\eqref{gsansatz1}
and \eqref{gsansatz2} divide the single-hole Hilbert space into four
sectors. The VMC simulation and analytic analysis have shown \cite{chen2019single}, consistent
with the ED and DMRG results, that besides $S_{z}=1/2$ for $|\Psi_{\mathrm{G}}\rangle_{1,2}$
and $S_{z}=-1/2$ for $|\Psi_{\mathrm{G}}\rangle_{3,4}$, each of
them acquires an additional double degeneracies with nontrivial orbital
momenta $L_{z}=\pm1$, respectively, for a bipartite lattice of $N=2M\times2M$.

The phase shift factors $e^{\mp i\hat{\Omega}_{i}}$ in Eqs.~(\ref{gsansatz1})
and (\ref{gsansatz2}) characterize the mutual entanglement between
the hole and spins in the AF background $|\phi_{0}\rangle$ to precisely
keep track of the phase-string structure. Explicitly, we can make
a choice $\hat{\Omega}_{i}=\sum_{l}\mathrm{Im}\ln(z_{i}-z_{l})n_{l}^{\downarrow}$, as in Eq.~\eqref{phaseO}
for a 2D system. It represents a mutual duality transformation, which
cannot be perturbatively obtained based on a bare hole state created
by $c_{i\sigma}$ on $|\phi_{0}\rangle$. With the nontrivial many-body
effect encoded in $e^{\mp i\hat{\Omega}_{i}}$, the resulting $\varphi_{1,2,3,4}\left(i\right)$
are consequently treated as the single-hole wave functions in the
VMC calculation, which have turned out to be a very good approximation
\cite{chen2019single}.

Given the single-particle nature, 
nevertheless, $\varphi_{1,2,3,4}\left(i\right)$ are not simply Bloch
wave functions. By a symmetry analysis, we show some intrinsic topological
relations among them in the following.

The anti-unitary time reversal symmetry $\mathcal{T}$ flips spin
direction by 
\begin{equation}
\mathcal{T}c_{\uparrow}\mathcal{T}^{-1}=c_{\downarrow},\mathcal{T}c_{\downarrow}\mathcal{T}^{-1}=-c_{\uparrow},\mathcal{T}i\mathcal{T}^{-1}=-i~.\label{App:timerreveral}
\end{equation}
Then by noting the time-reversal symmetry in the ground states, one
has 
\begin{equation}
\mathcal{T}|\Psi_{G}\rangle_{1}=|\Psi_{G}\rangle_{3}~,
\end{equation}
which requires 
\begin{equation}
\varphi_{3}\left(i\right)=-\varphi_{1}^{\ast}\left(i\right)e^{i\sum_{l\not=i}\mathrm{Im}\ln\left(z_{i}-z_{l}\right)}~,
\end{equation}
with the opposite $S_{z}=\pm1/2$ and $L_{z}=\pm1$. Similarly, $\mathcal{T}$
transforms the state $|\Psi_{G}\rangle_{2}$ into $|\Psi_{G}\rangle_{4}$,
which results in 
\begin{equation}
\varphi_{4}\left(i\right)=-\varphi_{2}^{\ast}\left(i\right)e^{i\sum_{l\not=i}\mathrm{Im}\ln\left(z_{i}-z_{l}\right)}~.
\end{equation}

Furthermore, one may also define a unitary symmetry $\mathbb{Z}_{2}$
generated by the spin flip operation $\mathcal{R}$, which is defined
\begin{equation}
\mathcal{R}c_{\uparrow}\mathcal{R}^{-1}=c_{\downarrow},\quad\mathcal{R}c_{\downarrow}\mathcal{R}^{-1}=-c_{\uparrow}~,\label{App:spinflip}
\end{equation}
with no extra sign change in the imaginary unit $i$. Then, 
\begin{equation}
\mathcal{R}|\Psi_{\mathrm{G}}\rangle_{1}=|\Psi_{\mathrm{G}}\rangle_{4}
\end{equation}
leads to 
\begin{equation}
\varphi_{4}(i)=-e^{-i\sum_{l\not=i}\mathrm{Im}\ln\left(z_{i}-z_{l}\right)}\varphi_{1}(i)~,
\end{equation}
where $|\Psi_{\mathrm{G}}\rangle_{1}$ and $|\Psi_{\mathrm{G}}\rangle_{4}$
have the same $L_{z}=-1$ and opposite $S_{z}=\pm1/2$. Similarly,
$\mathcal{R}$ transforms the state $|\Psi_{\mathrm{G}}\rangle_{2}$
into $|\Psi_{\mathrm{G}}\rangle_{3}$. 
Table~\ref{App:Tab} lists the corresponding quantum numbers of the
degenerate ground states $|\Psi_{\mathrm{G}}\rangle_{1,2,3,4}$ and
the actions of time reversal $\mathcal{T}$ and spin flip symmetries
$\mathcal{R}$.

\begin{table}
\label{App:Tab} \caption{List of the quantum numbers of four degenerate ground states $|\Psi_{\mathrm{G}}\rangle_{1,2,3,4}$
in the pure single-hole doped $t$-$J$.
}
\centering %
\begin{tabular}{cccccc}
\hline 
 & Phase string factor  & $S_{z}$  & $L_{z}$  & $J_{z}$  & Relation with $|\Psi\rangle_{1}$ \tabularnewline
\hline 
$|\Psi_{\mathrm{G}}\rangle_{1}$  & $e^{-i\hat{\Omega}_{i}}$  & $+\frac{1}{2}$  & $+1$  & $+\frac{3}{2}$  & \tabularnewline
$|\Psi_{\mathrm{G}}\rangle_{2}$  & $e^{+i\hat{\Omega}_{i}}$  & $+\frac{1}{2}$  & $-1$  & $-\frac{1}{2}$  & $\mathcal{RT}|\Psi\rangle_{1}$ \tabularnewline
$|\Psi_{\mathrm{G}}\rangle_{3}$  & $e^{-i\hat{\Omega}_{i}}$  & $-\frac{1}{2}$  & $-1$  & $-\frac{3}{2}$  & $\mathcal{T}|\Psi\rangle_{1}$ \tabularnewline
$|\Psi_{\mathrm{G}}\rangle_{4}$  & $e^{+i\hat{\Omega}_{i}}$  & $-\frac{1}{2}$  & $+1$  & $+\frac{1}{2}$  & $\mathcal{R}|\Psi\rangle_{1}$ \tabularnewline
\hline 
\end{tabular}
\end{table}

\section{Rashba interaction}
\subsection{Total angular momentum $J_{z}$}
\label{App:AngMom} 

The presence of the Rashba interaction breaks
conservation of the angular momentum $L_{z}$ and total spin $S_{z}$.
However, the total angular momentum $J_{z}=L_{z}+S_{z}$ remains as
a good quantum number.
Here we give a proof. We aim to prove
$UH_{t\text{-}J\text{-R}}U^{\dag}=H_{t\text{-}J\text{-R}}$ for $t$-$J$-R
Hamiltonian with $U=\exp(-i\frac{\pi}{2}J_{z})$ generated by $J_{z}$.
For Rashba SOC term $c_{i\uparrow}^{\dag}c_{i+\hat{x}\downarrow}^ {}$
in Eq.~\eqref{RashbaH}, 
\begin{align}
Uc_{i\uparrow}^{\dag}c_{i+\hat{x}\downarrow}^ {}U^{\dag} & =e^{-i\frac{\pi}{2}(L_{z}+S_{z})}c_{i\uparrow}^{\dag}c_{i+\hat{x}\downarrow}^ {}e^{i\frac{\pi}{2}(L_{z}+S_{z})}\nonumber \\
 & =e^{-i\frac{\pi}{2}S_{z}}c_{\hat{R}(i)\uparrow}^{\dag}c_{\hat{R}(i+\hat{x})\downarrow}^ {}e^{+i\frac{\pi}{2}S_{z}}\nonumber \\
 & =-ic_{\hat{R}(i)\uparrow}^{\dag}c_{\hat{R}(i)+\hat{y}\downarrow}^ {},~\label{App:term1}
\end{align}
where $\hat{R}(i)$ is the site by rotating site $i$ counterclockwisely
90 degrees. Similarly, for term $c_{i\uparrow}^{\dag}c_{i+\hat{y}\downarrow}^ {}$,
we have 
\begin{equation}
Uc_{i\uparrow}^{\dag}c_{i+\hat{y}\downarrow}^ {}U^{\dag}=-ic_{\hat{R}(i)\uparrow}^{\dag}c_{\hat{R}(i)-\hat{x}\downarrow}^{}~.\label{App:term2}
\end{equation}
Combination of Eqs.~\eqref{App:term1} and \eqref{App:term2} demonstrates
the invariance of Rashba interaction under $U$,
\begin{equation}
UH_{\text{R}}U^{\dag}=H_{\text{R}}~.
\end{equation}
The hopping and superexchange terms obviously conserve the total momentum
$J_{z}$. Therefore, we prove that total momentum $J_{z}$ is a good
quantum number for the $t$-$J$-R model.

\begin{figure}
\centering 
\includegraphics[scale=0.5]{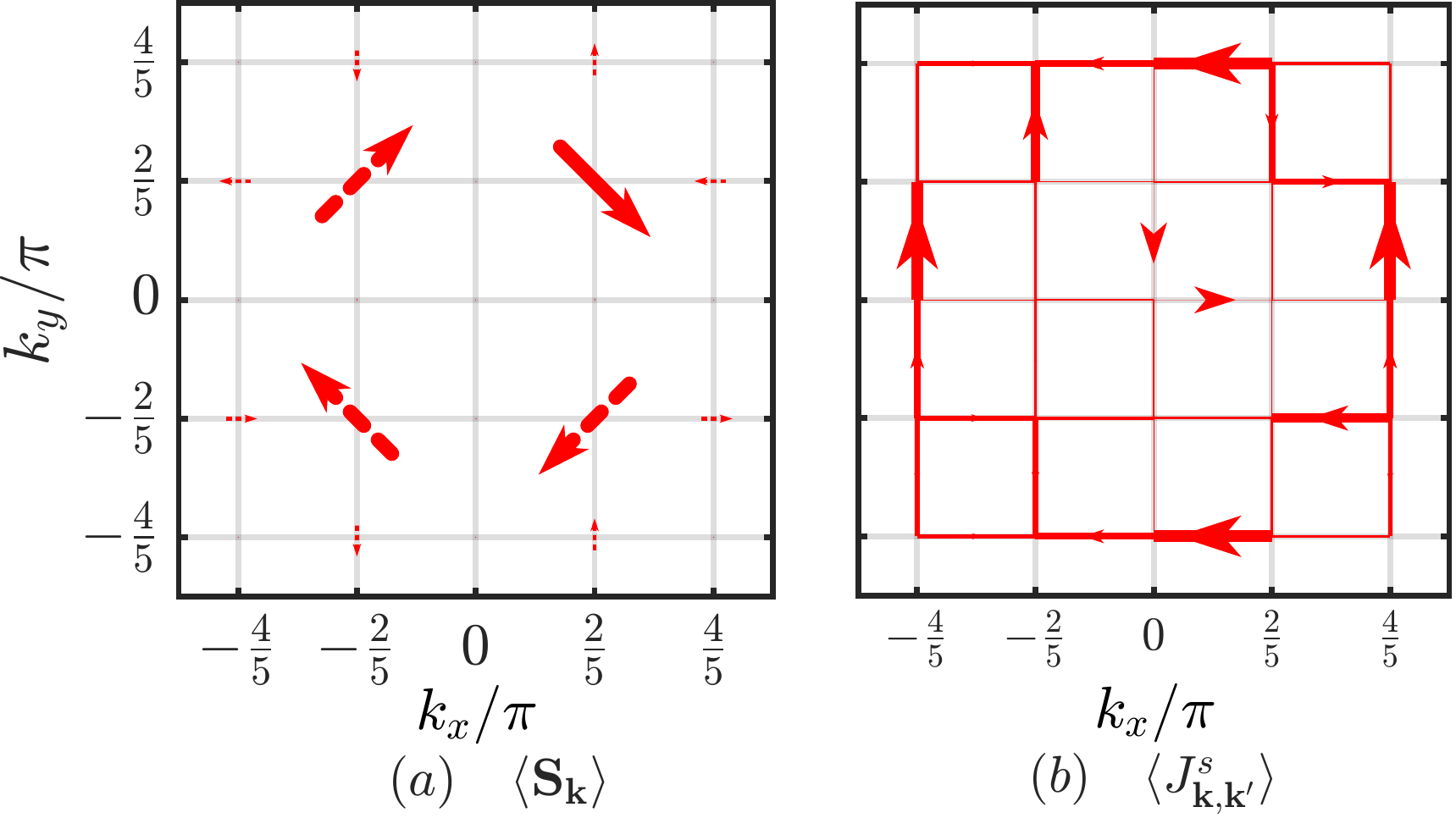} 
\caption{Two-component structure of the one-hole ground states of the $t$-$J$
model with a weak SOC ($\lambda=0.006J$) obtained by ED under PBC
in a $4\times4$ lattice: (a) The total spin $S=1/2$ is locked with
four total momenta at $\mathbf{k}_{0}=(\pm\pi/2,\pm\pi/2)$; (b) Corresponding
to one of the $\mathbf{k}_{0}$'s in (a) (the full arrow), a \textquotedblleft spin
current\textquotedblright{} pattern of $\langle J^s_{\mathbf k,\mathbf k^\prime }\rangle$
in Eq.~\eqref{eqn:Jk}
indicates a hidden many-body effect (see text). }
\label{Fig:MutualCurr} 
\end{figure}


\subsection{ED under periodic boundary condition}

\label{App:PBC}

We present the ED result under the periodic boundary condition (PBC)
in the following. Generally, without SOC, there are eight-fold degenerate
ground states distinguished by momenta $\mathbf{k}_{0}=\left(\pm\pi/2,\pm\pi/2\right)$
with total $S^{z}=\pm1/2$ in the pure one-hole state \cite{zheng2018hidden}.
The Rashba SOC will lift the spin degeneracy such that the spin $S=1/2$
will be locked with the four momenta $\mathbf{k}_{0}$ to give rise
to a spin texture of $\left\langle \mathbf{S}_{\mathbf{k}}\right\rangle $
lying in the $x$-$y$ plane as illustrated in Fig.~\ref{Fig:MutualCurr}(a),
which is obtained by ED with a very small SOC strength $\lambda=0.006J$
(see below). Here, corresponding to a ground state with a given total
momentum, say, $\mathbf{k}_0=\left(\pi/2,\pi/2\right)$, 
$\left\langle \mathbf{S}_{\mathbf{k}}\right\rangle  $
is mainly concentrated at $\mathbf{k}_0$ as indicated by the solid
arrow in Fig.~\ref{Fig:MutualCurr}(a) similar to a \emph{free} quasiparticle.

However, different from the free quasiparticle picture, in the present
ground state, a nontrivial ``spin current'' pattern is also found
in the momentum space as shown in Fig.~\ref{Fig:MutualCurr}(b),
which is defined by the correlator 
\begin{equation}
\langle J_{\mathbf{k},\mathbf{k}^{\prime}}^{s}\rangle\equiv i\left\langle S_{\mathbf{k}}^{+}S_{\mathbf{k}^{\prime}}^{-}-S_{\mathbf{k}^{\prime}}^{+}S_{\mathbf{k}}^{-}\right\rangle ~.
\label{eqn:Jk}
\end{equation}
Here $\langle J_{\mathbf{k},\mathbf{k}^{\prime}}^{s}\rangle$ illustrates
correlations between spins at the neighboring momenta (with $\mathbf{k}\neq\mathbf{k}'$),
which is present even in the absence of the SOC. The nontrivial spin
current pattern in Fig.~\ref{Fig:MutualCurr}(b) only disappears
in the half-filling due to fully frozen charge degree of freedom and the
absence of phase string effect. 
Hence, $\langle J_{\mathbf{k},\mathbf{k}^{\prime}}^{s}\rangle\neq0$
represents an incoherent component of the one-hole ground state, which
manifests as a broaden momentum distribution of the single hole as
a partial momentum can be carried away by the backflow spin current
such that the single-particle momentum $\mathbf{k}$ is no longer
conserved \cite{zheng2018hidden}. This represents the so-called incoherent
component violating the Landau's one-to-one correspondence between the
momentum $\mathbf{k}$ of the hole and the total momentum $\mathbf{k}_{0}$.
The latter should be always conserved under PBC due to the translational
invariance of the total system including all spins and the hole. However,
due to the spin current, a partial momentum can be continuously transferred
to the spin background from the hole or \emph{vice versa} to result
in a non-Landau quasiparticle state, which is essentially a many-body
effect \cite{zheng2018hidden}.

\begin{figure}[t]
\begin{centering}
\includegraphics[scale=0.33]{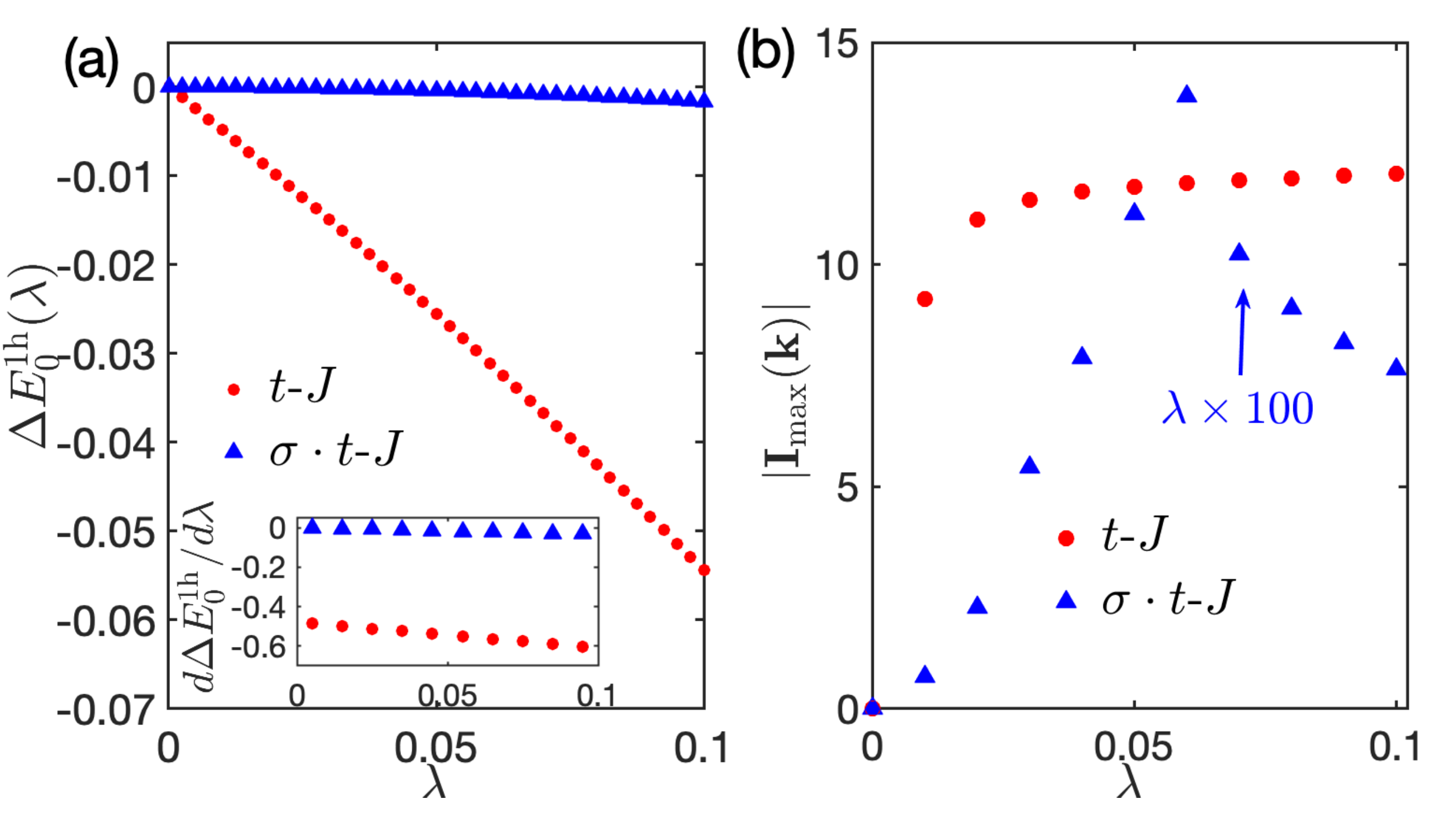} 
\par\end{centering}
\caption{ (a) Ground state energy shift $\Delta E_{0}^{\mathrm{1h}}$ 
vs. the SOC strength $\lambda$ for 
the $t$-$J$ (red dot) and the $\sigma\cdot $$t$-$J$ (blue triangle)
models, respectively. Inset: the corresponding first derivatives of
$\Delta E_{0}^{\mathrm{1h}}$; (b) Intensity of the spin-polarized
single-particle spectral function $|\mathbf{I}_{\mathrm{max}}(\mathbf{k})|$
at $\mathbf{k}=(\pi/2,\pi/2)$ as a function of $\lambda$. Here ED
results are obtained in a $4\times4$ lattice under PBC.}
\label{Fig:tjPBCEnergyInten} 
\end{figure}

Next we examine the quantitative role of SOC. Figure~\ref{Fig:tjPBCEnergyInten}(a)
shows the \emph{change} of the ground-state energy $\Delta E_{0}^{\mathrm{1h}}(\lambda)\equiv E_{0}^{\mathrm{1h}}(\lambda)-E_{0}^{\mathrm{1h}}(0)$
as a function of $\lambda$. For the $t$-$J$ model, the linear-$\lambda$
dependence (red line) and a \emph{finite} value of its first derivative
over $\lambda$ at $\lambda\rightarrow0$ [cf. the inset of Fig.~\ref{Fig:tjPBCEnergyInten}(a)]
strongly suggest that the spin texture in Fig.~\ref{Fig:MutualCurr}
at a very small $\lambda=0.006J$ is \emph{not} completely new as
induced by the SOC, but should rather exist already in the pure $t$-$J$
model to give rise to the first-order perturbative contribution. 
Namely, a weak $\lambda$ here mainly selects and picks up a particular
linear recombination state from the pure one-hole ($2\times4=8$ fold)
degenerate ground states in forming the spin texture in Figs.~\ref{Fig:MutualCurr}(a)
and \ref{Fig:MutualCurr}(b).

Indeed, by turning off the phase string effect or the spin current
structure in the $t$-$J$ model to result in a Landau-like quasiparticle
in the $\sigma\cdot $$t$-$J$ model, the change in ground state energy
due to SOC becomes significantly weakened as comparatively shown by
Fig.~\ref{Fig:tjPBCEnergyInten}(a) and the inset (blue triangles).
There is no more linear-$\lambda$ term. As a matter of fact, to make
the spin texture amplitude comparable, the SOC strength $\lambda$
has to be enhanced by two orders of magnitude in $\sigma\cdot $$t$-$J$
model as illustrated in Fig.~\ref{Fig:tjPBCEnergyInten}(b). Here
the intensity $|\mathbf{I}_{\max}\left(\mathbf{k}\right)|$ of the
spin-polarized spectral function [defined in Eq.~(\ref{IntenVect})
in Appendix \ref{Sec:Spectfun}] is shown in Fig.~\ref{Fig:tjPBCEnergyInten}(b)
at $\mathbf{k}=\left(\pi/2,\pi/2\right)$.

\section{Variational Monte Carlo procedure}
\label{APP:VMC}

\subsection{Variational ground state at half-filling}

\label{APP:VMChf} 
At half filling, both $t$-$J$ and $\sigma\cdot $$t$-$J$
models are reduced to a pure spin-$1/2$ Heisenberg system. At present,
the best variational ground state is the so-called 
Liang-Doucot-Anderson bosonic ``resonating valence bond'' (RVB)
state \cite{liang1988some} as 
\begin{equation}
|\phi_{0}\rangle=\sum_{\upsilon}\omega_{\upsilon}|\upsilon\rangle~,
\end{equation}
where each pair of spins in a dimer covering configuration $|\upsilon\rangle$
from different sublattices 
\begin{equation}
|\upsilon\rangle=\sum_{\{\sigma\}}(\prod_{(i,j)\in\upsilon}\epsilon_{\sigma_{i}\sigma_{j}})c_{1\sigma_{1}}^{\dagger}\cdots c_{N\sigma_{N}}^{\dagger}|0\rangle~,
\end{equation}
forms a singlet pairing forced by a Levi-Civita symbol $\epsilon_{\sigma_{i}\sigma_{j}}$
or, equivalently the Marshall sign \cite{auerbachbook}. During the
calculations, the non-negative amplitude $\omega_{\upsilon}$ can be
factorized as $\omega_{\upsilon}=\prod_{(ij)\in\upsilon}h_{ij}$,
where $h_{ij}$ is a non-negative function depending on sites $i$
and $j$. To obtain a ground state for AF spin Heisenberg model, we
have to optimize all factors $h_{ij}$. Conversely, with predetermined
$h_{ij}$, one can artificially tune the background spin correlations
$\xi$. For example, in Fig.~\ref{Fig:fintedoping}, we choose a
form of $h_{ij}$, 
\begin{equation}
h_{ij}=r_{ij}^{-\alpha}~,\label{App:hij}
\end{equation}
where $r_{ij}$ being the Manhattan distance $|x_{i}-x_{j}|+|y_{i}-y_{j}|$
with $(x_{i},y_{i})$ being a coordinate of site $i$. The spin correlation
length $\xi$ gradually decreases as the parameter $\alpha$ in Eq.~\eqref{App:hij}
increases. For example, a state with $\alpha=2$ shows an 
AF long range order 
 $\xi\gg L$, as shown in Fig.~\ref{Fig:schematicplot}, while a short range ordered 
 state with $\xi\sim 2.7$ can be obtained by setting $\alpha=4$.

\subsection{ Variational wave functions for the $t$-$J$-R model}

\label{APP:VMCtJR} Based on the ground states in Eqs.~(\ref{gsansatz1})
and (\ref{gsansatz2}), the wave function $\varphi_{1,2,3,4}\left(i\right)$
can be determined by VMC. For example, $\varphi_{1,2}\left(i\right)$
will be obtained by diagonalizing the following effective Hamiltonian
$H_{\mathrm{eff}}$ as the wave function for the composite quasiparticle
$\tilde{c}_{i\downarrow}$ defined in Eq.~(\ref{eqn:ctilde}): 
\begin{equation}
H_{\mathrm{eff}}=-\sum_{\langle i,j\rangle}\tilde{t}_{ij}^{\downarrow}h_{i}^{\dagger}h_{j}+h.c.~,\label{App:selfHam}
\end{equation}
where $\tilde{t}_{ij}^{\downarrow}$ represents an effective hopping
integral 
\begin{equation}
\tilde{t}_{ij}^{\downarrow}=t\sum_{\sigma}\langle\phi_{0}|\tilde{c}_{j\downarrow}^{\dagger}c_{j\sigma}c_{i\sigma}^{\dagger}\tilde{c}_{i\downarrow}|\phi_{0}\rangle~.\label{t_eff}
\end{equation}
Here $\tilde{t}_{ij}^{\downarrow}$ generally takes a complex value
due to the phase shift factor of a twisted particle $\tilde{c}_{i\downarrow}\equiv e^{\mp i\hat{\Omega}_{i}}{c}_{i\downarrow}$.
Thus, it is a Harper-Hofstadter-like Hamiltonian with a non-uniform
flux in Eq.~\eqref{App:selfHam} that breaks the translational symmetry
for $\varphi_{1,2}\left(i\right)$. Such an emergent flux can induce
exotic consequences beyond the simple tight binding model. In such
an effective Hamiltonian, $h_{i}^{\dagger}$ creates a `twisted' hole 
with the wave function $\varphi_{1,2}\left(i\right)$.
Two sectors are available, denoted
as $|\Psi\rangle_{1,2}$ respectively in Eq.~(\ref{gsansatz1}), which
correctly reproduce two ground states $|\Psi_{\text{G}}\rangle_{1,2}$
with the angular momentum $L_{z}=\pm1$. 
For $S_{z}=-\frac{1}{2}$, similar
constructions $|\Psi\rangle_{3,4}$ in Eq.~\eqref{gsansatz2} lead
to two ground states $|\Psi_{\text{G}}\rangle_{3,4}$ with orbital
momentum $L_{z}=\mp1$.
Indeed, excited states can also be determined by Eq.~\eqref{App:selfHam}
as one may choose excited wave functions $\varphi_{1,2,3,4}$ determined by
Eq.~\eqref{t_eff} or artificially tune the spin correlation in $\vert \phi_0\rangle$ [cf. Appendix~\ref{APP:VMChf}].

Now let us turn on the SOC with $\lambda\neq0$. As noted above, $L_{z}=\pm1$
and $S^{z}=\pm1/2$ are no longer the good quantum numbers, but the
total angular momentum $J_{z}=L_{z}+S_{z}$ remains. Then one may
reconstruct the one-hole ansatz states in terms of the linear combinations
between $|\Psi_{\mathrm{G}}\rangle_{1,2,3,4}$ and their excited states.
From the symmetry discussed above, Rashba interaction will lift the
degeneracy between $|\Psi_{\mathrm{G}}\rangle_{1,3}$ and $|\Psi_{\mathrm{G}}\rangle_{2,4}$
with different $J_{z}$'s. A ground state ansatz for the $t$-$J$-R
model should come from a linear combination, i.e., 
\begin{align}
\!\!\!\!\!|\Psi_{\mathrm{R}}\rangle_{\mathrm{umdm}} & =|\Psi\rangle_{1}+|\Psi\rangle_{3}\nonumber \\
 & =\sum_{i}[\varphi_{\downarrow}(i)e^{-i\hat{\Omega}_{i}}c_{i\downarrow}+\varphi_{\uparrow}(i)e^{-i\hat{\Omega}_{i}}c_{i\uparrow}]|\phi_{0}\rangle,\label{vmcwf13}
\end{align}
or between $|\Psi\rangle_{2}$ and $|\Psi\rangle_{4}$, 
\begin{align}
\!\!\!|\Psi_{\mathrm{R}}\rangle_{\mathrm{updp}} & =|\Psi\rangle_{2}+|\Psi\rangle_{4}\nonumber \\
 & =\sum_{i}[\varphi_{\downarrow}(i)e^{+i\hat{\Omega}_{i}}c_{i\downarrow}+\varphi_{\uparrow}(i)e^{+i\hat{\Omega}_{i}}c_{i\uparrow}]|\phi_{0}\rangle,\label{vmcwf24}
\end{align}
where hole wave functions $\varphi_{\uparrow}$ and $\varphi_{\downarrow}$
are variational parameters and the subscripts `umdm' or `updp' represent
chirality of the built-in phase string factor $e^{\pm i\hat{\Omega}_{i}}$
with `m' for `$-$' and `p' for `$+$' in the front of the removed
$\uparrow$-spin (`u') or $\downarrow$-spin (`d'). 
The true ground states 
are variationally proved to be $\vert \Psi_\mathrm{R}\rangle_\mathrm{umdm}$ or Eq.~\eqref{wfansatz-so},
and its two-fold
degeneracy is protected by a time reversal symmetry $\mathcal{T}$
in Eq.~\eqref{App:timerreveral}.

Parallel to the ansatz in Eqs.~(\ref{vmcwf13}) and (\ref{vmcwf24}),
another two recombinations $|\Psi_{\mathrm{R}}\rangle_{\text{umdp}}$
and $|\Psi_{\mathrm{R}}\rangle_{\mathrm{updm}}$ will generate excited states.
Therefore, we have four sectors of variational wave functions $|\Psi_{\mathrm{R}}\rangle_{\mathrm{umdm}}$,
$|\Psi_{\mathrm{R}}\rangle_{\text{updp}}$, $|\Psi_{\mathrm{R}}\rangle_{\mathrm{updm}}$
and $|\Psi_{\mathrm{R}}\rangle_{\text{umdp}}$. Fig.~\ref{Fig:EnergyEDVMC}(b)
shows energies of ground states with $J_{z}=\pm3/2$ and the first
excited states with $J_{z}=\pm1/2$ over the SOC strength $\lambda$
via VMC as compared with Fig.~\ref{Fig:EnergyEDVMC}(a) via ED.

Small Rashba interaction allows perturbative analysis on ground state
wave functions. Conservation of $J_{z}$ requires the ground states
for the $t$-$J$-R model to be composed of one of ground states 
and one excited state with $L_{z}=\pm2$ for the $t$-$J$ model. A
sufficiently small SOC strength $\lambda$ will select one chiral
spin pattern with total momentum $J_{z}=\frac{3}{2}$ or $-\frac{3}{2}$,
such that $|\Psi_{\mathrm{R}}\rangle_{\mathrm{updp}}$ produces ground
states. If we choose $|\Psi\rangle_{2}$ to be $|\Psi_{\mathrm{G}}\rangle_{2}$
with $J_{z}=\frac{3}{2}$. Then $|\Psi\rangle_{4}$ must have a quantum number
$L_{z}=2$ with $J_{z}=2-\frac{1}{2}=\frac{3}{2}$, which is the first
excited state of the $t$-$J$ model. If we start with $|\Psi_{\text{G}}\rangle_{4}$ as a choice for $|\Psi\rangle_{4}$, 
then $|\Psi\rangle_{2}$ is the
first excited state of $t$-$J$ model with $L_{z}=-2$. 
The two ground states can be transformed to each other by time-reversal symmetry $\mathcal T$ in Eq.~\eqref{App:timerreveral}. 
Technically, as proved by previous ED results \cite{zheng2018hidden}, non-degeneracy
between $|\Psi_{\mathrm{G}}\rangle_{1,2,3,4}$ and first excited states accidentally
originates from the open boundary condition, which leads to dramatically
large energy gain induced by SOC as compared to the $\sigma\cdot $$t$-$J$-R
model in Fig.~\ref{Fig:EnergyEDVMC}(a) and (b).

\section{Spin-polarized spectral function}

\label{Sec:Spectfun}

The spin-polarized angle-resolved photoemission spectroscopy (ARPES)
can be used to detect the spin texture of a quasiparticle (hole) excitation.
The normal spectral function for a single quasihole of spin $\sigma$
is defined by 
\begin{align}
\mathcal{A}_{\sigma}(\omega,\mathbf{k}) & =\sum_{n}\langle\Psi_{\mathrm{G}}|c_{\mathbf{k}\sigma}^{\dag}|n\rangle\langle n|c_{\mathbf{k}\sigma}^ {}|\Psi_{\mathrm{G}}\rangle\delta(\omega-E_{n}+E_{\text{G}})\nonumber \\
 & =\frac{1}{\pi}\mathrm{Im}\sum_{n}\frac{\langle\Psi_{\mathrm{G}}|c_{\mathbf{k}\sigma}^{\dag}|n\rangle\langle n|c_{\mathbf{k}\sigma}^ {}|\Psi_{\mathrm{G}}\rangle}{\omega-E_{n}+E_{\text{G}}-i\eta}~,\label{SpectrumFunc}
\end{align}
where $|\Psi_{\mathrm{G}}\rangle$ is the ground state with energy
$E_{\text{G}}$ and $|n\rangle$ denotes an eigenstate of energy $E_{n}$
with an extra hole created by $c_{\mathbf{k},\sigma}^ {}$ on $|\Psi_{\mathrm{G}}\rangle$.
In Eq.~(\ref{SpectrumFunc}), $\eta$ is a broadening introduced
to represent energy resolution. Based on $\mathcal{A}_{\sigma}(\omega,\mathbf{k})$,
we may further extract information on spin polarization, for example,
by focusing on the intensity of the scattered electrons that are parallel
or perpendicular to the momentum $\mathbf{k}$. One may define 
\begin{align}
c_{\mathbf{k},\sigma_{\perp}}^{\dag} & =\frac{1}{\sqrt{2}}\left(c_{\mathbf{k}\uparrow}^{\dag}+ie^{i\theta_{\mathbf{k}}}c_{\mathbf{k}\downarrow}^{\dag}\right)~,\\
c_{\mathbf{k},\bar{\sigma}_{\perp}}^{\dag} & =\frac{1}{\sqrt{2}}\left(c_{\mathbf{k}\uparrow}^{\dag}-ie^{i\theta_{\mathbf{k}}}c_{\mathbf{k}\downarrow}^{\dag}\right)~,
\end{align}
to create holes respectively with spin polarization at $\mathbf{S}_{\sigma_{\perp}}\mathbf{=}\left(\cos\theta_{\mathbf{k}},\sin\theta_{\mathbf{k}},0\right)$
and $\mathbf{S}_{\bar{\sigma}_{\perp}}\mathbf{=}-\mathbf{S}_{\sigma_{\perp}}$
perpendicular to momentum $\mathbf{k}$ lying in the $x$-$y$ plane with
$e^{i\theta_{\mathbf{k}}}=\frac{k_{x}+ik_{y}}{\left\vert \mathbf{k}\right\vert }$.
Similarly, the two operators 
\begin{align}
c_{\mathbf{k},\sigma_{\parallel}}^{\dag} & =\frac{1}{\sqrt{2}}\left(c_{\mathbf{k}\uparrow}^{\dag}+e^{i\theta_{\mathbf{k}}}c_{\mathbf{k}\downarrow}^{\dag}\right)~,\\
c_{\mathbf{k},\bar{\sigma}_{\parallel}}^{\dag} & =\frac{1}{\sqrt{2}}\left(c_{\mathbf{k}\uparrow}^{\dag}-e^{i\theta_{\mathbf{k}}}c_{\mathbf{k}\downarrow}^{\dag}\right)~,
\end{align}
create holes with spin polarization along $\mathbf{S}_{\sigma_{\parallel}}\mathbf{=}\left(\sin\theta_{\mathbf{k}},-\cos\theta_{\mathbf{k}},0\right)$
and $\mathbf{S}_{\bar{\sigma}_{\parallel}}\mathbf{=}-\mathbf{S}_{\sigma_{\parallel}}$,
parallel to momentum $\mathbf{k}$. Then the spin polarization $I_{\bot}\left(\omega,\mathbf{k}\right)$
at the direction perpendicular to momentum $\mathbf{k}$ takes the
form 
\begin{align}
I_{\bot}(\omega,\mathbf{k}) & =\mathcal{A}_{\sigma_{\bot}}(\omega,\mathbf{k})-\mathcal{A}_{\bar{\sigma}_{\bot}}(\omega,\mathbf{k})~,\label{SpinPolarizationPerp}
\end{align}
and $I_{\parallel}\left(\omega,\mathbf{k}\right)$ at direction parallel
to momentum $\mathbf{k}$: 
\begin{align}
I_{\parallel}(\omega,\mathbf{k}) & =\mathcal{A}_{\sigma_{\parallel}}(\omega,\mathbf{k})-\mathcal{A}_{\bar{\sigma}_{\parallel}}(\omega,\mathbf{k})~.\label{SpinPolarizationPara}
\end{align}
Finally one may define an intensity vector $\mathbf{I}_{\mathrm{max}}(\mathbf{k})$
\begin{equation}
\mathbf{I}_{\mathrm{max}}(\mathbf{k})=I_{\parallel}(\omega_{c},\mathbf{k})\mathbf{e}_{\mathbf{k},\parallel}+I_{\perp}(\omega_{c},\mathbf{k})\mathbf{e}_{\mathbf{k},\perp}~,\label{IntenVect}
\end{equation}
where $\mathbf{e}_{\mathbf{k},\parallel}$ ($\mathbf{e}_{\mathbf{k},\perp}$)
is the unit vector parallel (perpendicular) to momentum $\mathbf{k}$
and $\omega_{c}$ denotes the frequency at which $\sqrt{I_{\perp}(\omega,\mathbf{k})^{2}+I_{\parallel}(\omega,\mathbf{k})^{2}}$
reaches its maximum value $\vert\mathbf{I}_{\mathrm{max}}(\mathbf{k})\vert$.
The quantity $\mathbf{I}_{\mathrm{max}}(\mathbf{k})$ as probed by
the spin-polarized ARPES experiment can effectively characterize the
spin texture structure in the momentum space, away from the Fermi
energy and deep inside the Brillouin zone.

\section{Fragility of spin texture in a Fermi Liquid}
\label{App:FL} 

In a Fermi liquid, the spin texture induced by Rashba
SOC is fragile at momentum away from the Fermi surface. In this appendix,
we give a brief discussion. 

The Landau's Fermi liquid theory asserts one-to-one correspondence
between a free fermi gas system and an interacting one. The interaction
as scattering between quasiparticles manifests as a self energy $\Sigma(\omega,\mathbf{k})$
in the Green's function of electrons $c_\sigma$
\begin{align}
\mathcal G_\sigma(\omega,\mathbf k) & =\frac{1}{G_{0}^{-1}(\omega,\mathbf{k})-\Sigma(\omega,\mathbf{k})}\nonumber \\
 & =\frac{1}{\omega-[\epsilon_{0}(\mathbf{k})-\mu+\mathrm{Re}\Sigma(\omega,\mathbf{k})]-i\Gamma(\omega,\mathbf{k})}~,
 \label{eq:elegreen}
\end{align}
where $G_{0}(\omega,\mathbf{k})=[\omega-(\epsilon_{0}(\mathbf{k})-\mu)]^{-1}$
denotes the propagator for the non-interacting Hamiltonian $H_{0}$,
which generally involves both nearest and next-nearest neighbor hopping,
$t$ and $t^{\prime}$ with a dispersion relation $\epsilon_{0}(\mathbf{k})$,
\begin{equation}
\epsilon_{0}(\mathbf{k})=-2t(\cos k_{x}+\cos k_{y})+4t^{\prime}\cos k_{x}\cos k_{y}~.
\end{equation}
$\mu$ is a chemical potential which determines the doping concentration.
The imaginary part $\Gamma(\omega,\mathbf{k})$ in Eq.~(\ref{eq:elegreen})
of the self energy $\Sigma(\omega,\mathbf{k})$ dominates the lifetime
of a quasiparticle excitation. A significant consequence is that to the leading order, $\Gamma(\omega,\mathbf{k})$
depends on the second power of the exciting energy, 
\begin{equation}
\Gamma(\omega,\mathbf{k})=c\omega^{2}~,
\end{equation}
with a coefficient $c$ determined by quasiparticles' scatterings.

A sufficiently weak Rashba SOC interaction, on a reasonable assumption,
exerts no influence on the self-energy. In other words, we shall introduce
eigenmodes $d_{\mathbf{k}\pm}^\dagger$ to diagonalize the free part $H_{0}+H_{R}$,
\begin{equation}
    d_{\mathbf{k}\pm}^{\dagger} = \frac{1}{\sqrt{2}}(c_{\mathbf{k}\uparrow}^{\dagger}\pm e^{i\varphi_{{\bf k}}}c_{\mathbf{k}\downarrow}^{\dagger})~,
\end{equation}
which, consequently, modify the Green's functions 
\begin{align}
    \mathcal G_{+}(\omega,\mathbf k) & =\frac{1}{G_{0}^{-1}(\omega,\mathbf{k})-\Delta_{\mathbf{k}}-i\Gamma(\omega, \mathbf{k})}~,\\
    \mathcal G_{-}(\omega, \mathbf k) & =\frac{1}{G_{0}^{-1}(\omega,\mathbf{k})+\Delta_{\mathbf{k}}-i\Gamma(\omega,\mathbf{k})}~.
\end{align}
Here $\Delta_{\mathbf{k}}=2\lambda\sqrt{\sin^{2}k_{x}+\sin^{2}k_{y}}$
is the energy splitting induced by SOC in $H_{R}$, and 
$e^{i\varphi_{{\bf k}}}=\frac{\sin k_{x}+i\sin k_{y}}{\sqrt{\sin^{2}k_{x}+\sin^{2}k_{y}}}$.

Follow the procedure presented in Sec.~\ref{Sec:Spectfun} and we
can obtain the spin polarization perpendicular to the momentum $\mathbf{k}$
\begin{align}
    I_{\perp}(\omega_{c},\mathbf{k}) & = \mathcal A_{\sigma_\perp} (\omega_c, \mathbf k)-\mathcal A_{\bar{\sigma}_\perp} (\omega_c, \mathbf k)\notag \\ 
    &= 2\cos(\theta_{\mathbf{k}}-\varphi_{\mathbf{k}})\frac{2\Delta_{\mathbf{k}}}{c\epsilon_{{\bf k}}^{2}(c\epsilon_{{\bf k}}^{2}+\Delta_{\mathbf{k}})}~,
    \label{IntenFL}
\end{align}
where $\omega_{c}$ is defined in Eq.~(\ref{IntenVect}) to relate
to the maximal spectral function $\mathbf{I}_{\mathrm{max}}(\mathbf{k})$ and $e^{i\theta_{\mathbf{k}}}=\frac{k_{x}+ik_{y}}{\left\vert \mathbf{k}\right\vert }$.
Since $\Delta_{\mathbf{k}}$ is very small, a quasiparticle with momentum
away Fermi surface towards $\Gamma$ point makes a vanishing contribution,
which is depicted in Fig.~\ref{Fig:SpinTextureFL} with parameters
$c=0.005,\lambda=0.01J$ and $t^\prime=0.2t$.

\end{document}